\documentclass{aa}   
\usepackage{graphicx}
\usepackage{txfonts,textcomp}
\usepackage{amsmath}   
\usepackage{gensymb}
\usepackage[normalem]{ulem}
\usepackage{comment}
\usepackage{array}
\usepackage{float}
\usepackage{lipsum}
\usepackage{adjustbox}
\usepackage{threeparttable}

\usepackage[colorlinks =TRUE, linkcolor = blue,
            urlcolor  = blue,
            citecolor = blue,
            anchorcolor = blue]{hyperref}
\begin{document}

\title{IXPE and \textit{VLT}/FORS2 polarimetry challenge the Seyfert-1.9 classification of MCG-05-23-16}

\titlerunning{Multi-wavelength polarimetry reclassifies MCG-05-23-16 as a type-1 AGN}
  
   \author{Frédéric Marin\inst{1}  
           \and
           Daniele Tagliacozzo\inst{2}  
           \and
           Francesco Ursini\inst{2}    
           \and
           Damien Hutsemékers\inst{3}  
           \and
           Mitsuru Kokubo\inst{4} 
           \and
           Thibault Barnouin\inst{1}   
           \and   
           Andrea Gnarini\inst{2,5}   
           \and   
           Alessandro Leonardo Lai\inst{2}   
           \and  
           Jiří Svoboda\inst{6}               
           \and
           Stefano Bianchi\inst{2}  
           \and
           Vittoria Elvezia Gianolli\inst{7}  
           \and
           Ephraim Gau\inst{8,9}  
           \and
           Kun Hu\inst{8,9}  
           \and
           Henric Krawczynski\inst{8,9,10}  
           \and
           W. Peter Maksym\inst{5}  
           \and
           Andrea Marinucci\inst{11}
           \and
           Herman Marshall\inst{12}  
           \and
           Giorgio Matt\inst{2}  
           \and
           Riccardo Middei\inst{13}  
           \and
           Pierre-Olivier Petrucci\inst{14} 
           \and
           Simonetta Puccetti\inst{15}
           \and
           Nicole Rodriguez\inst{8,9,10}  
           \and
           Roberto Serafinelli\inst{16,13}  
           \and
           Francesco Tombesi\inst{17,18}  
          }           

   \institute{Universit\'e de Strasbourg, CNRS, Observatoire Astronomique de Strasbourg, UMR 7550, 11 rue de l'universit\'e, 67000 Strasbourg, France\\
             \email{frederic.marin@astro.unistra.fr}             \and            
                 Dipartimento di Matematica e Fisica, Università degli Studi Roma Tre, via della Vasca Navale 84, 00146 Roma, Italy
             \and            
                 Institut d’Astrophysique et de G\'eophysique, Universit\'e de Li\`ege, All\'ee du 6 Ao\^ut 19c, B5c, 4000 Li\`ege, Belgium
             \and            
                 National Astronomical Observatory of Japan, National Institutes of Natural Sciences, 2-21-1 Osawa, Mitaka, Tokyo 181-8588, Japan
            \and
                NASA Marshall Space Flight Center, Huntsville, AL 35812, USA
            \and
                Astronomical Institute of the Czech Academy of Sciences, Bo\v{c}n\'i II 1401/1, 14100 Praha 4, Czech Republic 
            \and
                Department of Physics and Astronomy, Clemson University, Kinard Lab of Physics, Clemson, SC 29634, USA
            \and
                Department of Physics, Washington University, St. Louis, MO 63130, USA
            \and
                McDonnell Center for the Space Sciences, Washington University, St. Louis, MO 63130, USA
            \and
                Center for Quantum Leaps, Washington University, St. Louis, MO 63130, USA
            \and
                ASI - Agenzia Spaziale Italiana, Via del Politecnico snc, 00133 Roma, Italy
            \and
                MIT Kavli Institute for Astrophysics and Space Research, Massachusetts Institute of Technology, 77 Massachusetts Avenue, Cambridge, MA 02139, USA
            \and
                INAF Osservatorio Astronomico di Roma, Via Frascati 33, 00078 Monte Porzio Catone, RM, Italy
            \and
                Université Grenoble Alpes, CNRS, IPAG, 38000 Grenoble, France
            \and
                Space Science Data Center, Agenzia Spaziale Italiana, Via del Politecnico snc, 00133 Roma, Italy
            \and
                Instituto de Estudios Astrofísicos, Facultad de Ingeniería y Ciencias, Universidad Diego Portales, Avenida Ejército Libertador 441, Santiago, Chile
            \and
                Dipartimento di Fisica, Universitá degli Studi di Roma “Tor Vergata,” Via della Ricerca Scientifica 1, 00133 Roma, Italy
            \and
                Istituto Nazionale di Fisica Nucleare, Sezione di Roma “Tor Vergata,” Via della Ricerca Scientifica 1, 00133 Roma, Italy
             }

   \date{Received November 19, 2025; accepted January 20, 2026}

   \abstract
{We report the third observation of the Seyfert-1.9 active galactic nucleus (AGN) MCG-05-23-16 with the Imaging X-ray Polarimetry Explorer (\textit{IXPE}), together with optical spectro-polarimetry obtained at the Very Large Telescope (VLT), and combined with archival near-ultraviolet, optical and near-infrared polarimetric data. No X-ray polarization was detected in the 2–8 keV band, with a 99\% confidence upper limit of $\leq$2.9\%, further reduced to $\leq$2.5\% when combined with the two past IXPE observations of the same target. Monte Carlo simulations suggest that equatorial coronal models are disfavored if the AGN is indeed a type-1.9/2 AGN, but coronae coplanar to the accretion disk remain consistent if the source is less inclined than previously assumed. \textit{VLT}/FORS2 data reveal a typical type-2 spectrum in total flux, a broad H$\alpha$ line in polarized flux, and strongly wavelength dependent polarization degree and angle, rotating by nearly 70$^\circ$ across the optical band. Comparison with historical measurements confirms long-term stability of the polarization spectrum and a $\sim$90$^\circ$ rotation in the near-ultraviolet. Interpreting the multi-wavelength polarization relative to the AGN ionization axis indicates that the main obscurer is not a compact circumnuclear torus, but a distant kpc-scale dust lane crossing the galaxy. This result implies that MCG-05-23-16 is in fact a type-1 AGN seen through foreground dust. The low  X-ray column density becomes consistent with the absence of polarization, provided that the nuclear inclination is low.}

   \keywords{galaxies: active – galaxies: individual: MCG-05-23-16 - galaxies: Seyfert – polarization – X-rays: galaxies}

%

    \maketitle

\section{Introduction}
\label{Introduction}

X-ray polarimetry has proven to be a powerful tool to probe the energy generation mechanisms at the heart of spatially unresolved sources. In particular, from microquasars in the hard state \citep[see, e.g.,][]{Krawczynski2022} to active galactic nuclei (AGNs) that are little or not obscured \citep{Gianolli2023}, high energy polarimetric measurements obtained with the Imaging X-ray Polarimetry Explorer (\textit{IXPE}, \citealt{Weisskopf2022}) have demonstrated that the region responsible for the emission of X-rays, a hot electron plasma situated near the accretion disk, is likely extended along the disk plane rather than being compact and situated along the disk symmetry axis, as it has long been considered in a widely-used lamp-post geometry \citep{Miniutti2004}. Two arguments support this conclusion: a high degree of linear polarization (several percent) and a polarization angle parallel to the axis traced by the radio structure/jets, indicating that the polarization emerges from scattering in a non-spherical equatorial region. By contrast, a more symmetric geometry (e.g., spherical) would reduce the polarization degree, and any other location (along the polar axis of the object, for example) would lead to a different polarization angle \citep{Ursini2022}.

NGC~4151 was the first radio-quiet, pole-on object in which X-ray polarimetry demonstrated that the X-ray corona is likely extended along the equatorial plane \citep{Gianolli2023,Gianolli2024}. IC~4329A, another unobscured AGN, seems to corroborate such findings but the results are less significant from a statistical point-of-view \citep{Ingram2023}. However, X-ray polarimetric data obtained by \textit{IXPE} are inconclusive so far for NGC~2110 \citep{Chakraborty2025} and for MCG-05-23-16. The last AGN is particularly interesting because, despite two pointings (May and October 2022), totalizing 1,128~Ms -- the largest exposure time for a radio-quiet, pole-on object --, only an upper limit of 3.2\% at 99\% confidence level has been inferred \citep{Marinucci2022,Tagliacozzo2023}. The polarization angle remains unconstrained. It is unsettling because MCG-05-23-16 is, among the unobscured (N$_{\rm H}$ $\sim$ 10$^{22}$ cm$^{-2}$, \citealt{Mattson2004}) AGNs observed by \textit{IXPE}, the one with the greatest estimated inclination (from 33$^\circ$ to 65$^\circ$, depending on the method, see \citealt{Turner1998,Reeves2007,Fuller2016,Serafinelli2023}), which should lead to high X-ray polarization levels. 

MCG-05-23-16 is a type-1.9 AGN, meaning that its optical broad emission lines are undetected in total flux. Only broad infrared emission lines (such as Pa$\beta$ emission) are detected, due to the decreasing impact of dust extinction in the infrared \citep{Goodrich1994}. Broad lines do exist in the optical spectrum of MCG-05-23-16, but they only appear in polarized light, thanks to scattering off the polar winds material \citep{Lumsden2004}. It hints for a large nuclear inclination and thus a large scattering-induced polarization too. But, despite its brightness (F$_{\rm 2-10~keV}$ = 7 -- 10 $\times$ 10$^{-11}$ erg~cm$^{-2}$~s$^{-1}$, \citealt{Mattson2004}), no X-ray polarization was detected from MCG-05-23-16. It either means that the source is much less inclined that was previously estimated or that its X-ray corona sustains an intrinsically lower polarization fraction, pointing to a different geometrical/physical configuration than all other unobscured AGNs and microquasars in the hard state observed by \textit{IXPE}.

To try to solve this issue, we conducted a third observation of MCG-05-23-16 with \textit{IXPE}  to combine the data from 2022 together with those obtained in 2025. The latter are supplemented by a new XMM-Newton observation to assess the spectral shape stability and disentangle the contributions to the polarization from different spectral components.

\section{X-ray observations and data reduction}
\label{Observation}

\textit{IXPE} observed MCG-05-23-16 on two previous occasions: in May 2022 \citep{Marinucci2022} and November 2022 \citep{Tagliacozzo2023}. The \textit{IXPE} observation, performed simultaneously with \textit{XMM-Newton} (for the first observation only) and \textit{NuSTAR} (for both campaigns), had a net exposure time of 486~ks and 642~ks, respectively. Both datasets, with updated response matrices, are also used in this work. The third pointing, which constitutes the observation that we report here, happened between April 24 and May 9, 2025. However, one of the three detector units (DUs) of \textit{IXPE} -- DU2 more precisely -- suffered from an anomaly and was thus completely excluded from the present analysis. The remaining two DUs (DU1 and 3) combined a total exposure of 745~ks. Cleaned level 2 event files were firstly treated with the background rejection procedure described in \citet{DiMarco2023}, then generated following standard filtering criteria using the dedicated \texttt{ftools} tasks and the latest available calibration files from the \textit{IXPE} CALDB (20250225). The analysis followed the formalism described by \citet{Strohmayer2017}, using the weighted analysis method detailed in \citet{DiMarco2022}, with \texttt{stokes=Neff} set in \texttt{xselect}. The $I$, $Q$, and $U$ Stokes background spectra were extracted from source-centered annular regions with an internal radius of 150~arcsec and external radius of 300~arcsec. The source extraction regions were optimized to maximize the Signal-to-Noise Ratio (SNR) in the 2--8~keV band, following the procedure described by \citet{Piconcelli2004}, leading to a choice of 85~arcsec and 80~arcsec radii for the first and third DUs, respectively. These same radii, centered on the source, were applied to the $I$, $Q$ and $U$ spectra. After processing, the net exposure time is 740~ks. A constant energy binning of 0.2~keV was adopted for the $Q$ and $U$ Stokes spectra, and a minimum SNR of 3 was required in each bin of the intensity spectra. The background contribution during the 2025 observation amounted to 2.4\% for each DU, quite comparable to the 2022 values. 

\textit{XMM-Newton} observed MCG-05-23-16 on May 8, 2025, with an elapsed time of 87.2~ks using the EPIC CCD cameras: the pn \citep{Struder2001} and the two MOS \citep{Turner2001}, operated in small window and medium filter mode. Data from the MOS detectors were excluded from the analysis due to pile-up, while the pn camera data showed no significant pile-up, as confirmed by the \texttt{EPATPLOT} output. The extraction radii and the optimal time cuts for background flares were computed using SAS~22.1 \citep{Gabriel2004}, following the same SNR maximization procedure adopted for previous observations. The resulting optimal extraction radii for the source and background spectra were 40 and 50~arcsec, respectively. The net exposure time for the pn time-averaged spectrum after filtering was 58.4~ks.

The background-subtracted light curves for all observations are presented in appendices. 

\section{Results and analysis}
\label{Results}

\subsection{IXPE campaign}
\label{Results:2025}

\begin{figure*}
\includegraphics[width=8.8cm] {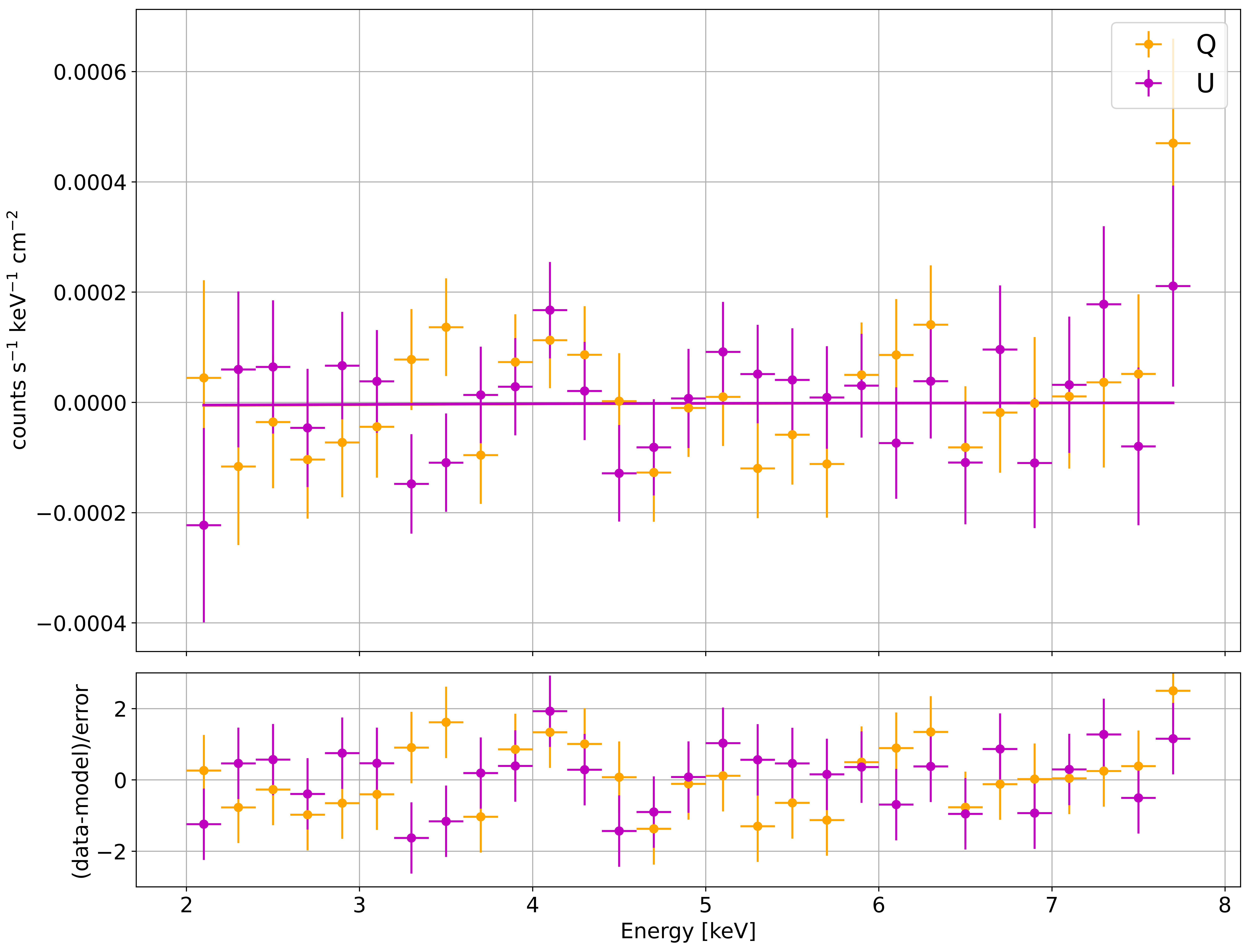}
\includegraphics[width=8.8cm] {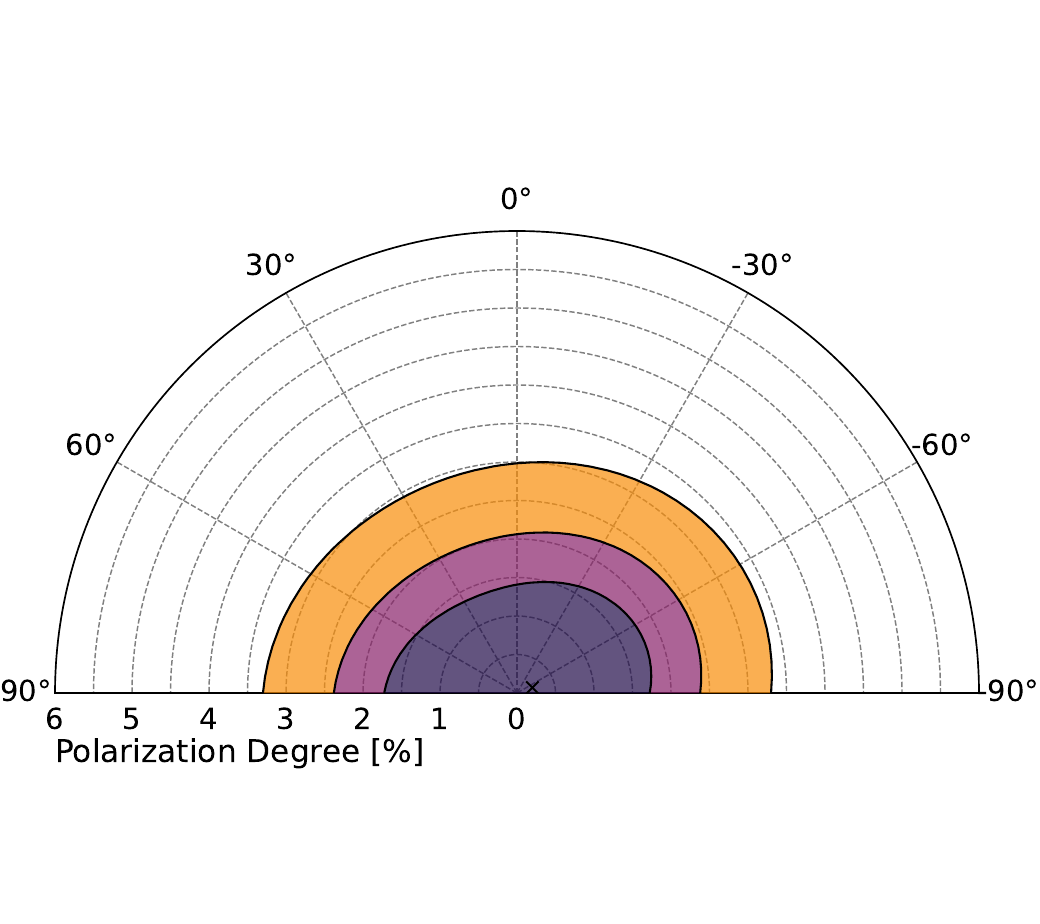}
\caption{\textit{Left panel:} \textit{IXPE} $Q$ (orange crosses) and $U$ (magenta crosses) grouped Stokes spectra of the 3rd \textit{IXPE} pointing (April 2025) of MCG-05-23-16 are shown with residuals, along with the corresponding best-fitting model. \textit{Right panel:} contour plot between the polarization degree $P$ and angle $\theta$, summed over the $2-8$ keV energy band of the 3rd \textit{IXPE} pointing. Purple, pink and orange regions correspond, respectively, to $68\%$, $90\%$ and $99\%$ confidence levels for two parameters of interest.}
\label{Fig:3rdobs}
\end{figure*}

First, we performed a preliminar analysis of the IXPE data alone, using \texttt{XSPEC} v12.13.1e \citep{Arnaud1996} and a baseline model composed of an absorbed power law convolved with a constant polarization kernel, as in \citet{Tagliacozzo2023}: $\texttt{const} \times \texttt{polconst} \times \texttt{tbabs} \times \texttt{ztbabs} \times \texttt{powerlaw}$. We applied this model to the third \textit{IXPE} observation of MCG-05-23-16 (April 2025; 740 ks), fitting simultaneously the $I$, $Q$, and $U$ spectra from the two available DUs over the 2--8~keV energy range. When relying solely on \textit{IXPE} data, more complex models are unnecessary, as evidenced by the reduced chi-square values consistently being close to one ($\chi^2$/d.o.f. = 415/405). The multiplicative constant is used to account for cross-calibration uncertainties among DU1 and DU3. The Galactic absorption along the line of sight was modeled with \texttt{tbabs}, assuming a column density of $N_{\mathrm{H,Gal}} = 7.8 \times 10^{20}~\mathrm{cm}^{-2}$ \citep{HI4PI2016}.
Freezing the column density at the redshift of the source ($z$ = 0.00849) to the value derived from a broad-band data analysis (see Sec. \ref{Results:Broadband}: $N_{H, z}=$1.48 $\times 10^{22}~\mathrm{cm}^{-2}$), we found a primary continuum spectral index of $\Gamma=1.84\pm0.01$. The left panel of Fig.~\ref{Fig:3rdobs} presents the $Q$ and $U$ spectra used in this initial analysis (including the model and the residuals), while the right panel shows the contour plot between the polarization degree $P$ and the polarization angle $\theta$ for the 3rd \textit{IXPE} observation.

In contrast to the 1st and 2nd observations of this source, the polarization degree is not constrained, even at $68\%$ confidence level. At $99\%$ confidence level, we find an upper limit of $P=3.0\%$. In addition, we do not find any significant variation of the polarization when evaluating it in smaller energy bins (i.e. we do not find any polarization energy trend). Compared to the 2022 observations (for which we recomputed the numbers using the latest calibration files and the same aforementioned baseline model), our 2025 campaign provides similar results, see Tab.~\ref{Tab:Data}. 
The polarization from the X-ray corona is consistent within 1$\sigma$ between the three epochs.

\begin{figure}
\includegraphics[width=8.8cm] {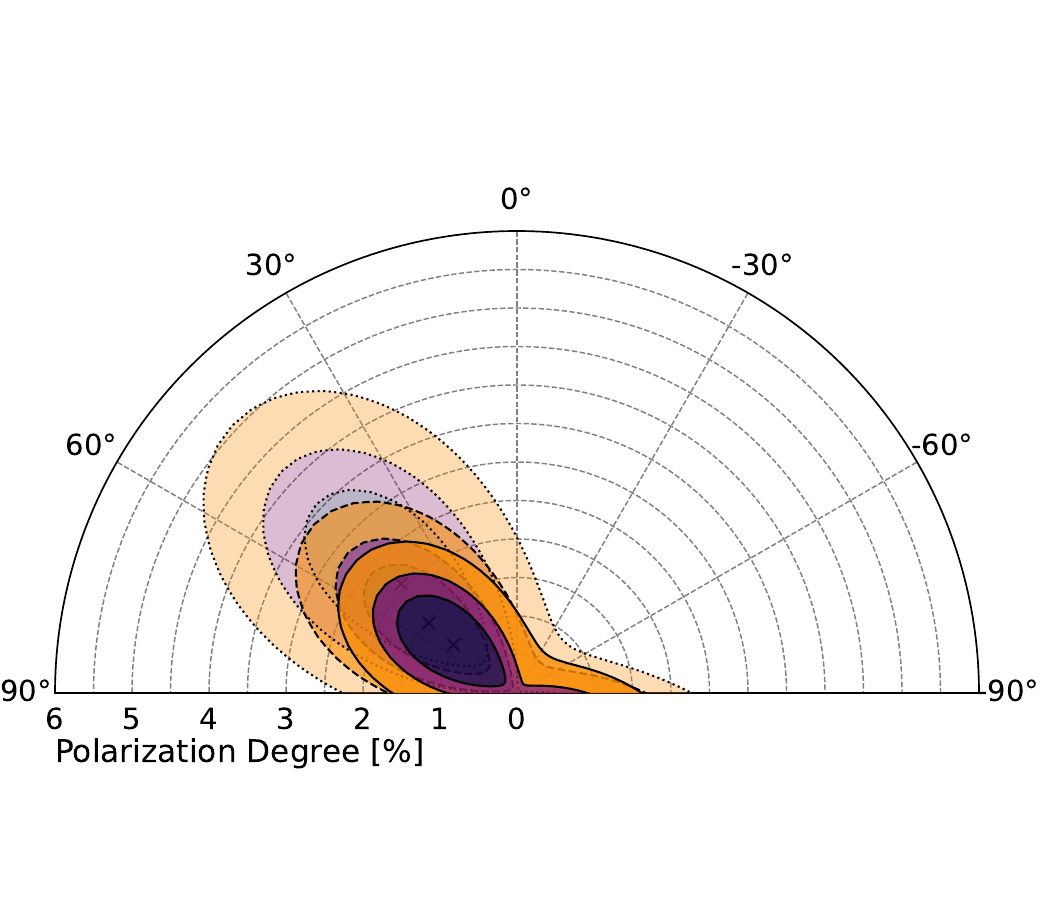}
\caption{$P$, $\theta$  contour plots of the May 2022 observation (dotted contours), May 2022 + November 2022 (dashed contours), May 2022 + November 2022 + April 2025 (solid contours). Contours colors are as in Fig.~\ref{Fig:3rdobs}.}
\label{Fig:IXPECombined}
\end{figure}

Next, we performed a combined analysis of the \textit{IXPE} $I$, $Q$, and $U$ spectra collected in May and November 2022, and in April 2025, using the same model and leaving the spectral indexes of the three pointings untied. As a result, we got the same value for the first two observations ($\Gamma_{1}=\Gamma_{2}=1.89\pm0.01$) and $\Gamma_{3}=1.81\pm0.01$ for the third. From this analysis, we obtained a polarization degree $P=1.0\pm0.6\%$ and a polarization angle $\theta=53\degree\pm18\degree$ at $68\%$ confidence level for one parameter of interest. This translates into a polarization degree upper limit (at $99\%$ confidence level) of 2.5\%. This represents a significant improvement with respect to the results obtained previously (see Tab.~\ref{Tab:Data} for the best-fit values of the polarization degree and angle obtained using only \textit{IXPE} dataset). Fig.~\ref{Fig:IXPECombined} shows the contour plots of the 1st observation, the 1st and the 2nd combined and the 1st, 2nd and 3rd combined. 

\begin{table}
\centering
\caption{X-ray polarimetry of MCG-05-23-16 using model-independent analyses.}
\begin{tabular}{lccc}
\hline
{Date} & {$P$ (\%)} & {$\theta$ (deg)} & {U.L. (\%)} \\ 
\hline
May 2022 & 2.2 $\pm$ 1.7 & 50 $\pm$ 24 & $\le$ 4.7 \\ 
Nov. 2022 & 1.1 $\pm$ 0.9 & 57 $\pm$ 27 & $\le$ 3.3 \\ 
Apr. 2025 & $\le1.1$ & -- & $\le3.0$ \\ 
All (2022+2025) & 1.0 $\pm$ 0.6  & 53 $\pm$ 18 & $\le$ 2.5 \\ 
\hline
\end{tabular}
\tablefoot{U.L. means "upper limit". The errors are shown at $68\%$ and the upper limits at $99\%$ confidence level for one parameter of interest.}
\label{Tab:Data}
\end{table}

\subsection{Broad-band spectral analysis}
\label{Results:Broadband}

To obtain a comprehensive understanding of the system and to obtain parameters relevant for a useful spectro-polarimetric analysis, we conducted a broad-band X-ray spectral analysis by combining the full set of \textit{IXPE} spectral ($I$ Stokes parameter) observations (May and November 2022, along with the 2025 campaign) with the 3–79 keV \textit{NuSTAR} spectra (May and November 2022) and the 2–10 keV \textit{XMM-Newton} spectra (May 2022 and April 2025). 

Following the methodology of \cite{Serafinelli2023}, we built the following model: 
\texttt{const} $\times$ \texttt{tbabs} $\times$ \texttt{ztbabs} [\texttt{zcutoffpl} + \texttt{vashift}(\texttt{relxill} + \texttt{xillver})], where we used \texttt{tbabs} to account for Galactic absorption (assuming a column density of $N_{\mathrm{H,Gal}} = 7.8 \times 10^{20}~\mathrm{cm}^{-2}$; \citealt{HI4PI2016}) and \texttt{ztbabs} to account for column density at the source's redshift. Also, we employed \texttt{zcutoffpl} to model the primary continuum, while \texttt{relxill} and \texttt{xillver} were used for relativistic (from an ionized disk with emissivity profile fixed to $\epsilon(r)\propto r^{-3}$) and distant (neutral; e.g. from the torus or the outer disk) reflection, respectively. These components share the photon index and cutoff energy of the primary continuum. The constant multiplicative component accounts for cross-calibration uncertainties among the various instruments (\textit{IXPE} DUs, \textit{NuSTAR} FPMA and FPMB, and \textit{XMM-Newton} EPIC-pn). As done in previous papers on this source \citep{Marinucci2022, Tagliacozzo2023,Serafinelli2023}, we added a \texttt{vashift} component -- which simply provides a shift in energy -- in order to deal with the energy of the narrow Fe K$\alpha$ line, which is inconsistent with being 6.4 keV in the host galaxy rest frame. This effect is found in the 2022 \textit{XMM}/pn (and not in the MOS) observation and in \textit{NuSTAR}/FPMA and FPMB (with an increasing deviation between the first and the second pointing) and it is likely due to calibration issues. Finally, to correct for known calibration issues in the May 2022 \textit{IXPE} observation, we followed the procedure of \citet{Marinucci2022} and \cite{Tagliacozzo2023}, adjusting the gain of the $I$ spectrum using the \texttt{gain fit} command in \texttt{XSPEC}. 

\begin{figure}[h!]
\includegraphics[width=9cm] {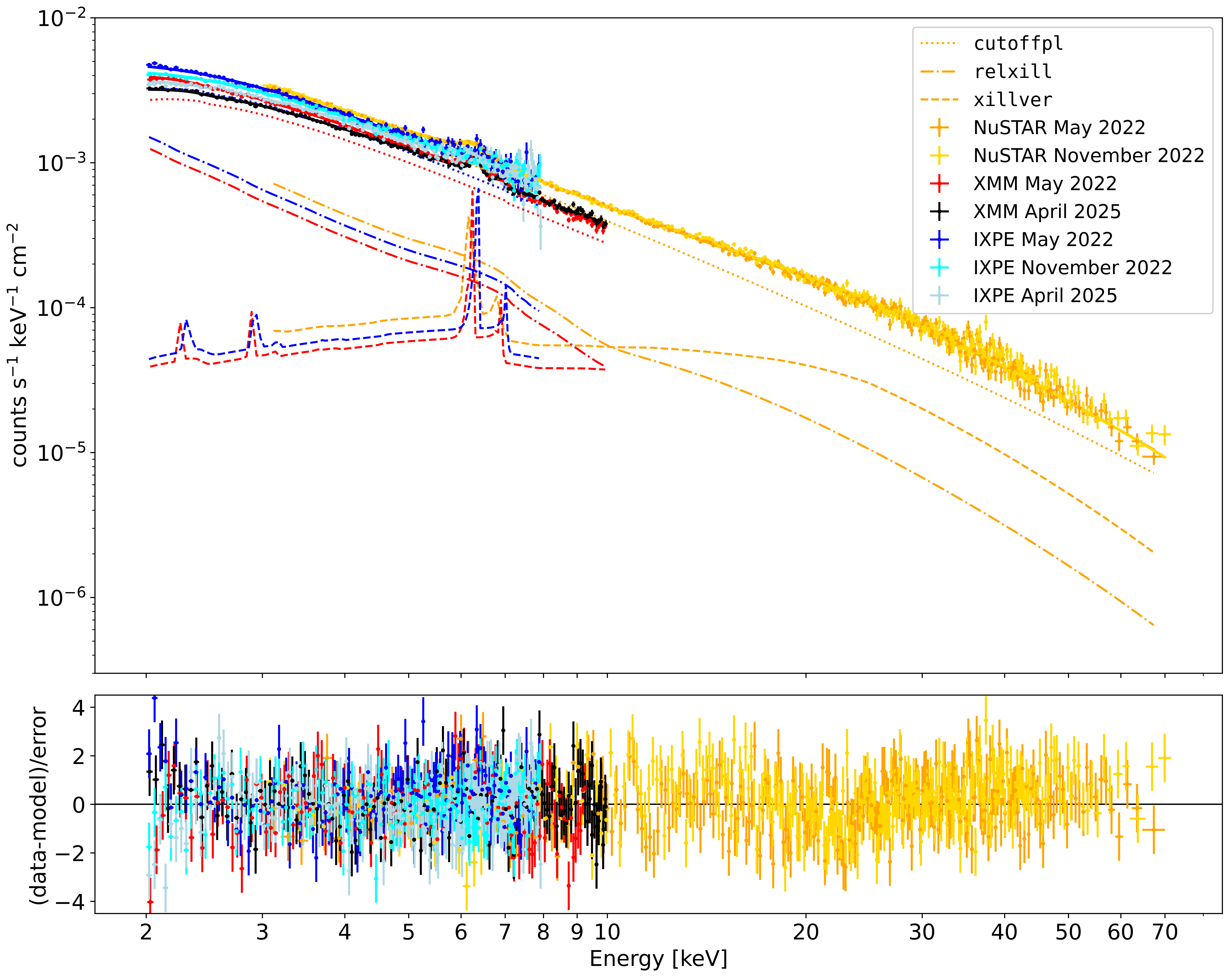}
\caption{\textit{NuSTAR}, \textit{XMM} and \textit{IXPE I} spectra, collected on May 2022, November 2022 and April 2025 of MCG-05-23-16, along with the best-fit model components for each instrument (only for the May 2022 campaign, in which the source was observed by the three instruments simultaneously) and the residuals for the broad-band (2--79~keV) spectral analysis in Sec. \ref{Results:Broadband}. \textit{NuSTAR} photon counting detector modules (FPMA and FPMB) of the same campaign have been grouped for visual clarity, as well as \textit{IXPE} DUs.}
\label{Fig:all_nopol}
\end{figure}

As an outcome, we notice a hardening of the spectrum itself between the 2022 and the 2025 campaigns: at $68\%$ confidence level we found $\Gamma_{\textnormal{2022}}=1.88^{+0.02}_{-0.01}$ and $\Gamma_{\textnormal{2025}}=1.75^{+0.03}_{-0.01}$. In fact, as in \cite{Tagliacozzo2023} and \cite{Serafinelli2023}, this parameter did not change significantly between the 1st and the 2nd observations. For this reason, we tied their spectral indexes together within this analysis. For the same reason -- plus the fact that we lack a \textit{NuSTAR} pointing within the 3rd campaign --, we let the primary continuum high energy cutoff free to vary, but we tied the values of all the three observations together, obtaining $E_C=116^{+6}_{-5}$ keV. While being initially tied between the three observations (resulting in a not fully satisfactory fit), we let all the absorption and reflection parameters free to vary. The fit significantly improved, but $\theta_{\textnormal{incl}}$ and R$_\textnormal{in}$ are not simultaneously constrained. This is likely due to the different sensitivity of the instruments and to the elevated number of parameters affecting the high-energy curvature. Therefore, as done by \cite{Serafinelli2023}, we fixed R$_\textnormal{in}$ to the assumed spin ISCO corresponding value, in order to retrieve $\theta_{\textnormal{incl}}$ (which is a fundamental parameter to determine useful polarization constraints). We obtained $\theta_{\textnormal{incl}}=35\degree\pm3\degree$. The fit is acceptable: $\chi^2$/d.o.f. = 2673/2396. In Fig.~\ref{Fig:all_nopol} we show the spectra used for this analysis along with the model components for each instrument (only for the May 2022 campaign, in which the source was observed by the three instruments simultaneously) and the residuals for the broad-band (2--79~keV) spectral analysis. Finally, in Tab.~\ref{Tab:all_nopol} we list the model parameters which have been retrieved in this analysis.

\begin{table}[h!]
\centering
\caption{Best-fitting parameters for the broad-band spectral analysis of MCG-05-23-16, using \textit{NuSTAR} (3--79~keV), \textit{XMM-Newton} (2--10~keV) and \textit{IXPE I} (2--8~keV) combined data set.}
\begin{tabular}{cc}
\hline
{Parameter}  & {Best fitting value} \\
\hline
$N_{\rm H,Gal}$  [cm$^{-2}$] & $7.8\times10^{20}$*  \\
$N_{\rm H,z}$ (May/November 2022) [cm$^{-2}$] & ($1.66_{-0.04}^{+0.08}$)$\times10^{22}$  \\
$N_{\rm H,z}$ (April 2025) [cm$^{-2}$] & ($1.48_{-0.03}^{+0.05}$)$\times10^{22}$  \\
\hline
\multicolumn{2}{c}{\texttt{zpl}} \\
 $\Gamma$ (May/November 2022)& $1.88^{+0.02}_{-0.01}$ \\
 $\Gamma$ (April 2025)& $1.75^{+0.02}_{-0.01}$ \\
 $E_{\rm C}$ [keV]& $116\pm5$  \\
 $N$ (May 2022)&($3.4\pm0.1$)$\times10^{-2}$  \\
 $N$ (November 2022)&($3.5\pm0.1$)$\times10^{-2}$  \\
 $N$ (April 2025)&($3.1\pm0.1$)$\times10^{-2}$  \\
 \hline
  \multicolumn{2}{c}{\texttt{relxill}} \\
  $\theta_{\textnormal{incl}}$ [deg]&$35\pm3$\\
  $a$&0.998*\\
  $R_{\textnormal{in}}$ [$R_{\rm G}$]&$1.24$*\\
  $\log{\xi}$ & $3.3\pm0.1$\\
   $N$ (May 2022)&$1.2\pm0.1$$\times10^{-4}$  \\
   $N$ (November 2022)&$7\pm1$$\times10^{-5}$  \\
 $N$ (April 2025)&($4\pm1$)$\times10^{-5}$  \\
  \hline
    \multicolumn{2}{c}{\texttt{xillver}} \\
  $A_{\textnormal{Fe}}$ &$0.9_{-0.06}^{+0.04}$\\
   $N$ (May 2022)&($3.3\pm0.2$)$\times10^{-4}$  \\
   $N$ (November 2022)&$(4.3\pm0.2)\times10^{-4}$  \\
 $N$ (April 2025)&$3.6_{-0.2}^{+0.3}$$\times10^{-4}$  \\
  \hline
  \multicolumn{2}{c}{\texttt{vashift} [km/s]}  \\
 \textit{NuSTAR} (May 2022) & $5.7_{-1.1}^{+0.9}\times 10^3$  \\
 \textit{NuSTAR} (November 2022) & $(4.8\pm0.7)\times 10^3$  \\
 \textit{XMM} (May 2022) & $(5.0\pm0.4)\times 10^3$  \\
 \textit{XMM} (April 2025) & //  \\
  \hline
    \multicolumn{2}{c}{Cross-calibration constants}  \\
 \textit{NuSTAR}/FPMB & $1.005\pm0.003$  \\
 \textit{XMM} & $0.697\pm0.002$  \\
 \textit{IXPE}/DU1 & $0.822\pm0.003$  \\
 \textit{IXPE}/DU2 & $0.856\pm0.003$  \\
 \textit{IXPE}/DU3 & $0.797\pm0.003$  \\
  \hline
   \multicolumn{2}{c}{ \textit{IXPE} $F_{2-10}$ [erg cm $^{-2}$s$^{-1}$]} \\ [0.4ex]
  May 2022& $(8.84-8.90)\times10^{-11}$\\
  November 2022& $(8.64-8.69)\times10^{-11}$\\
  April 2025 & $(8.58-8.64)\times10^{-11}$\\
  \hline
    \multicolumn{2}{c}{ \textit{IXPE} $L_{2-10}$ [erg s$^{-1}$]} \\ [0.4ex]
  May 2022& $(1.41-1.42)\times10^{43}$\\
  November 2022& $(1.38-1.39)\times10^{43}$\\
  April 2025 & $(1.44-1.45)\times10^{43}$\\
  \hline
 $R$ (May 2022)& 0.17\\
  $R$ (April 2025)& 0.16\\
  \hline
  $\chi^{2}$/dof& 2673/2396\\
  \hline
\end{tabular}
\begin{flushleft}{\textit{Note}: The errors at 68\% confidence level for one parameter of interest. Parameters which do not display the error and are followed by * symbol have been fixed during the fit.}
\end{flushleft} 
\label{Tab:all_nopol}
\end{table}

\begin{figure}[h!]
\includegraphics[width=9cm] {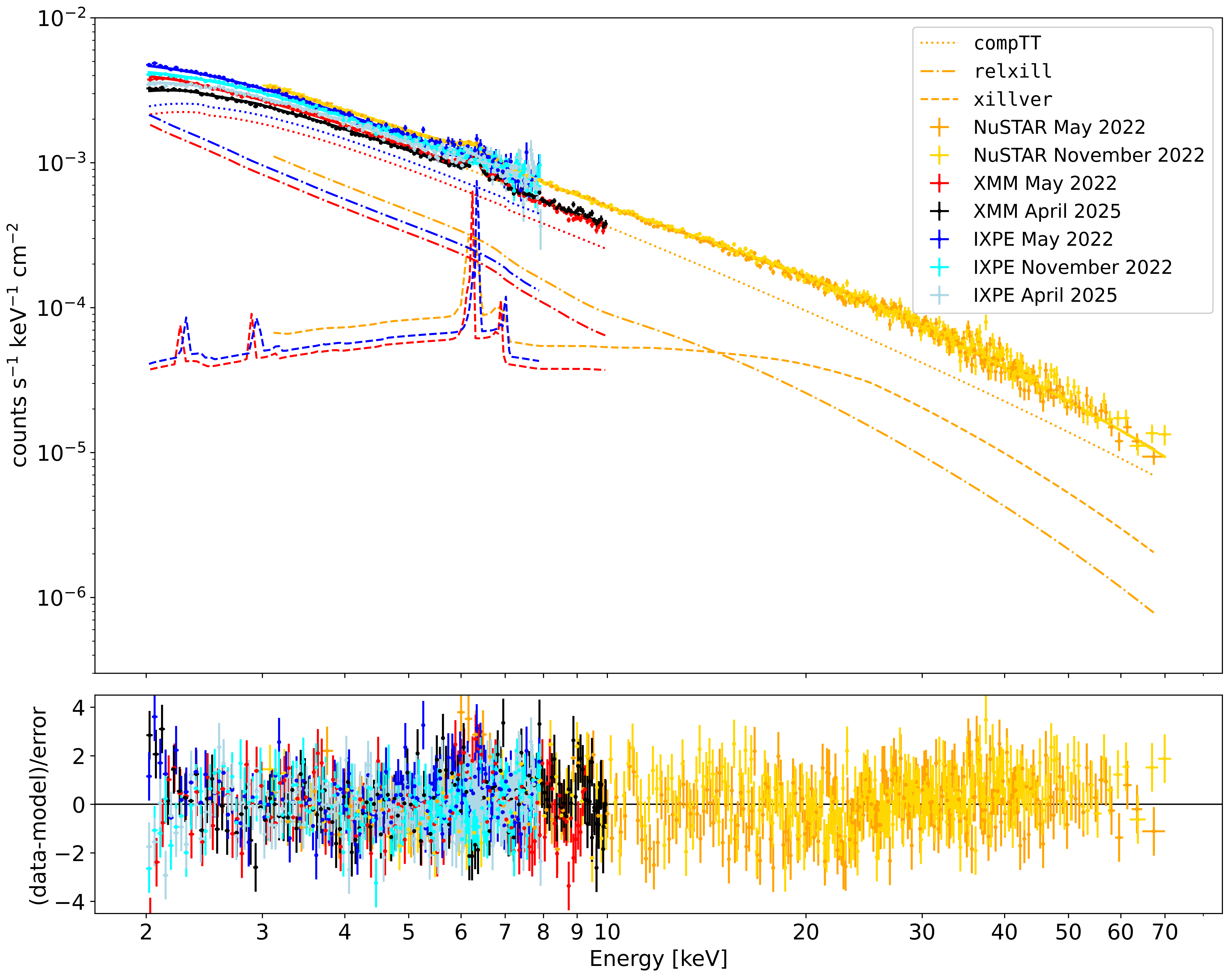}
\caption{Same as in Fig. \ref{Fig:all_nopol}, but with \texttt{compTT} (assuming a slab coronal configuration) instead of \texttt{cutoffpl}.}
\label{Fig:comptt_nopol}
\end{figure}

As the final step of this broad-band spectral analysis, we replaced \texttt{zcutoffpl} with the analytic Comptonization model \texttt{compTT} \citep{tita1994ApJ...434..570T}. We fixed the thermal photon temperature to 30 eV, which is appropriate for an AGN accretion disk and does not much influence the output spectrum \citep{Schneider2013}, and we fixed the reflection parameters to the values previously found. Assuming a slab geometry for the corona, we retrieved a coronal temperature $kT_e=36\pm2$ keV and a coronal optical depth $\tau=1.1\pm0.1$. Conversely, assuming a more spherical geometry, we retrieved $kT_e=33\pm2$ keV and $\tau=3.0\pm0.2$. All those measurements are consistent (within $3\sigma$) with previous inferences \citep{Balokovic2015,Marinucci2022,Serafinelli2023}. In Fig.~\ref{Fig:comptt_nopol}, we show the spectra used for this final spectral analysis along with the model components for each instrument (only for the May 2022 campaign, in which the source was observed by the three instruments simultaneously) and the residuals.

\subsection{Broad-band spectro-polarimetric analysis}
\label{Results:Broadband_pol}
Next, we assigned a distinct polarization signature to each model component using the convolutive polarization kernel \texttt{polconst} in \texttt{XSPEC}. Since the Fe K$\alpha$ line is expected to be intrinsically unpolarized \citep{Goosmann2011, Marin2018}, and the Compton reflection component contributes marginally to the \textit{IXPE} energy band \citep{Marin2018}, we initially fixed the polarization degree of both \texttt{relxill} and \texttt{xillver} to zero, leaving only the polarization properties of the primary continuum free to vary.

The overall fit is acceptable with $\chi^2$/d.o.f. = 3083/2879. Given the absence of significant residuals and the complexity of combining multiple instruments across different epochs, we attributed the residuals to cross-calibration systematics, as already mentioned in \citet{Tagliacozzo2023}. Also from this combined analysis, we only obtained an upper limit at $99\%$ confidence level on the primary continuum polarization, with $P<2.8\%$. As for the previous observations, although the polarization angle is unconstrained and that we lack a definitive polarization detection, our analysis suggests a statistical tendency for the polarization angle to be between 20 and 70$\degree$ (see left panel of Fig.~\ref{Fig:IXPECombined}). 

Next, we performed a series of tests by assigning different polarization degrees and angles to the reflection components (i.e. \texttt{relxill} and \texttt{xillver}) relative to the primary continuum component (\texttt{cutoffpl}) one. To assess the contribution of the Fe K$\alpha$ line (likely unpolarized) relative to the reflection continuum, we replaced the \texttt{xillver} component with \texttt{pexriv} \citep{mag1995MNRAS.273..837M} plus a gaussian line. The line has a small equivalent width of $68^{+16}_{-24}$ eV and a flux which is less than $1\%$ of the total reflection flux in the \textit{IXPE} energy range (i.e., 2--8~keV), resulting in no significant impact on the polarization constraints. Among the various tests we performed, which are summarized in Tab.~\ref{Tab:polresults} and similar to that of the work of \cite{Gianolli2023}, we forced the polarization of the reflection components to be perpendicular to the primary continuum one and we assumed physically motivated values of $P$ \citep{Marin2018}. In particular, we tested $P_{\texttt{xillver}}=P_{\texttt{relxill}}=10$, $20$ and $30\%$. Given that the overall 2--8~keV polarization is notably low, as indicated by our model-independent analysis (see Sect.~\ref{Results:2025}), assuming highly polarized reflected radiation naturally implies a correspondingly higher (assumed to be perpendicular to the reflection) polarization of the primary continuum, consistent with expectations for this scenario.

As a result, we obtained only an upper limit at the 99\% confidence level on $P_{\texttt{pl}}$ ($<$4\%) for $P_{\texttt{xillver}}=P_{\texttt{relxill}}=10\%$. When assuming a higher polarization for the reflected components, we measured $P_{\texttt{pl}}=(2.2\pm0.7)\%$ at the $68\%$ confidence level for $P_{\texttt{xillver}}=P_{\texttt{relxill}}=20\%$, and $P_{\texttt{pl}}=(3.7\pm0.7)\%$ for $P_{\texttt{xillver}}=P_{\texttt{relxill}}=30\%$, indicating a consistent increase of the inferred continuum polarization. These spectro-polarimetric tests suggest that the observed signal can be reproduced either by weakly polarized X-ray reflection ($\leq10\%$), implying a low intrinsic polarization of the primary continuum ($\leq4\%$), or by a scenario where more strongly polarized reflection ($\geq20\%$) is oriented nearly orthogonally to a few-percent polarized continuum, leading to partial cancellation of the net signal. The latter case would require rather specific scattering geometries.

\begin{table}
\centering
\caption{X-ray polarimetry of MCG-05-23-16. Model-dependent analyses.}
\begin{tabular}{ccc}
\hline
{$P/\theta_{\texttt{pl}}$} [$\%/\degree$] & {$P/\theta_{\texttt{relxill}}$} [$\%/\degree$] & {$P/\theta_{\texttt{xillver}}$} [$\%/\degree$]  \\ 
\hline
$\mathbf{<2.8/-}$ & $0/-$ & $0/-$ \\ 
$\mathbf{<8.4/-}$ & $\mathbf{<51/-}$ & $\mathbf{<73/-}$  \\ 
$\mathbf{<8.0/-}$ & $\mathbf{<51}/\perp\theta_{\texttt{pl}}$ & $\mathbf{<63}/\perp\theta_{\texttt{pl}}$   \\ 
$\mathbf{<2.6/-}$ & $\mathbf{<20}/\parallel\theta_{\texttt{pl}}$ & $\mathbf{<44}/\parallel\theta_{\texttt{pl}}$  \\ 
$\mathbf{<6.3/-}$ & $\mathbf{<48/-}$ & $\mathbf{<65}/\perp\theta_{\texttt{relxill}}$  \\ 
$\mathbf{<8.4/-}$ & $\mathbf{<51/-}$ & $\mathbf{<70}/\parallel\theta_{\texttt{relxill}}$  \\
$\mathbf{<3.5/-}$ & $\mathbf{<25}/\parallel\theta_{\texttt{pl}}$ & $\mathbf{<56}/\perp\theta_{\texttt{relxill}}$  \\
$\mathbf{<2.8/-}$ & $\mathbf{<44}/\perp\theta_{\texttt{xillver}}$ & $\mathbf{-}/\parallel\theta_{\texttt{pl}}$  \\
$\mathbf{<4/-}$ & $10/\perp\theta_{\texttt{pl}}$ & $10/\perp\theta_{\texttt{pl}}$  \\
$\mathbf{<2.2\pm0.7/-41\pm14}$ & $20/\perp\theta_{\texttt{pl}}$ & $20/\perp\theta_{\texttt{pl}}$  \\
$\mathbf{<3.7\pm0.7/-41\pm12}$ & $30/\perp\theta_{\texttt{pl}}$ & $30/\perp\theta_{\texttt{pl}}$  \\
\hline
\end{tabular}
\tablefoot{When the polarization degree is preceded by ``$<$'', the result is intended as an upper limit (at $99\%$ confidence level). The errors are shown at $68\%$ confidence level. $\perp$ means "perpendicular to" and $\parallel$ "parallel to". Parameters in bold font resulted from the fit, while parameters in normal font have been frozen before the fit procedure.}
\label{Tab:polresults}
\end{table}

\section{Comparison to simulations}
\label{Simulations}

In order to interpret our measurements, we explored the possible shape and location of the X-ray corona using Monte Carlo simulations. To do so, we used the code \texttt{MONK}, a Monte Carlo radiative transfer code dedicated to calculations of Comptonized spectra in the Kerr spacetime \citep{Zhang2019}. 

We simulated a series of X-ray coronae of different shapes and locations, following the approach of \cite{Ursini2022}, \cite{Gianolli2023}, \cite{Tagliacozzo2023}, \cite{Ingram2023}, and \cite{sudip2025ApJ...990...89C}.
We focused on the four most common and probable coronal configurations: 
\begin{itemize}
    \item a spherical lamppost corona. It is a spherical, compact and stationary source located along the disk axis. Its size depends on the supermassive black hole's spin. It tends to produce reflection-dominated spectra in AGN \citep{matt1991A&A...247...25M, marto1996MNRAS.282L..53M, pop1997A&A...326...99P, wilk2012MNRAS.424.1284W, ursi2020A&A...644A.132U, Ursini2022};
    \item conical outflows. They can be described as a conical mass located along the symmetry axis of the accretion disk. They are commonly associated with an aborted jet. According to this model, radio-quiet AGN host a central supermassive black hole that powers outflows and jets that may propagate only over short distances if the velocity of the ejected material is sub-relativistic and lower than the local escape velocity. Hence, the material would be able to travel only up to a certain distance before falling back and colliding with subsequent blobs. This would result in the dissipation of kinetic energy and in the production of high-energy emission in these objects \citep{pop1997A&A...326...99P, conerefId0, Ursini2022};
    \item a slab corona. In this scenario, the hot medium is assumed to be uniformly distributed above the disk. This geometry is induced by magnetic loops rising high above the disc plane and dissipating energy via reconnection. Here, the energy dissipation and electron heating occur over a large volume and the corona is assumed to be co-rotating with the Keplerian disk \citep{1979ApJ...231L.111L, Haard1991ApJ...380L..51H, belo2017ApJ...850..141B};
    \item a wedge corona. This configuration is similar to the slab one, but its height is increasing with the radius. In this scenario, the disk is thought to be truncated at a certain radius, while the corona represents some type of a "hot accretion flow", possibly extending to the innermost stable circular orbit (ISCO), see \citet{Esin_1997, sh2010ApJ...712..908S, Marcel2018, poutanen2018, Ursini2020, Tagliacozzo2025}.
\end{itemize}

We performed \texttt{MONK} simulations for a total of 24 parameter combinations. For the lamppost and slab cases, we simulated two spin configurations (null and maximum), each one for two disk radii ($25$ R$_{\textnormal{G}}$ and the ISCO, which is equal to $6$ R$_{\textnormal{G}}$ for the null spin case and $1.24$ R$_{\textnormal{G}}$ for the maximum spin case). For the wedge case, we simulated four different openings of the corona (15$\degree$, 30$\degree$, 45$\degree$ and 60$\degree$), for each of the spin and radii stated before. Finally, for the conical case, we relied on the results found by \cite{Ursini2022}, who simulated three different cases: maximum spin with height on the disk plane $d=3$ R$_{\textnormal{G}}$ and vertical thickness $t=10$ R$_{\textnormal{G}}$, null spin with height on the disk plane $d=5$ R$_{\textnormal{G}}$ and vertical thickness $t=15$ R$_{\textnormal{G}}$, and null spin with height on the disk plane $d=20$ R$_{\textnormal{G}}$ and vertical thickness $t=20$ R$_{\textnormal{G}}$. Since the \texttt{MONK} simulations in \cite{Ursini2022} have been performed with a coronal electron temperature of 25 keV, we adopted the same value, which is also consistent with our own values derived in Sect.~\ref{Results:Broadband} and the one measured by \cite{2015ApJ...800...62B}. However, considering the degeneracy between $kT_e$ and $\tau$ for a fixed spectral slope, a slight increase in $kT_e$ does not determine a significant variation of the expected polarization properties of the source, as found, e.g., by \cite{Tagliacozzo2025}. For each geometrical configuration, we then determined the optical depth that best reproduces the observed spectrum in the \textit{IXPE} bandpass (2–8~keV), by replacing the cutoff power-law component with the spectra generated by \texttt{MONK} in the best-fit model from Sect.~\ref{Results:Broadband}. For all the simulations we performed, we assumed a mass of the supermassive black hole of $M_{\rm BH}=2\times10^7M_{\odot}$ and an Eddington ratio of 0.1 \citep{po2012A&A...542A..83P}. Finally, we set the initial polarization (i.e. the polarization of the optical/ultraviolet radiation emitted by the accretion disk) as appropriate for a pure scattering, plane-parallel, semi-infinite atmosphere \citep{1960ratr.book.....C}.

As expected, we found polarization angles parallel to the accretion disk axis in the slab and wedge cases. Conversely, for the lamppost and conical cases, we find angles perpendicular to the accretion disk axis. The polarization degree is highest for the slab geometry, lowest for the lamppost and intermediate for the wedge and the conical coronas. In the lamppost scenario, $P$ is always lower than 1.5$\%$. For the conical case, the configuration with null spin, $d=20$ R$_{\textnormal{G}}$ and $t=20$ R$_{\textnormal{G}}$ results in the highest $P$ (up to 6$\%$). The slab corona cases with null spin results in lower $P$ (up to 7$\%$), while cases with maximum spin result in higher $P$ (up to 13$\%$). Finally, in the wedge scenario, already investigated by \cite{Tagliacozzo2025}, $P$ is maximum for lower opening angles and higher black hole spins.

\begin{figure*}
\includegraphics[width=6.375cm] {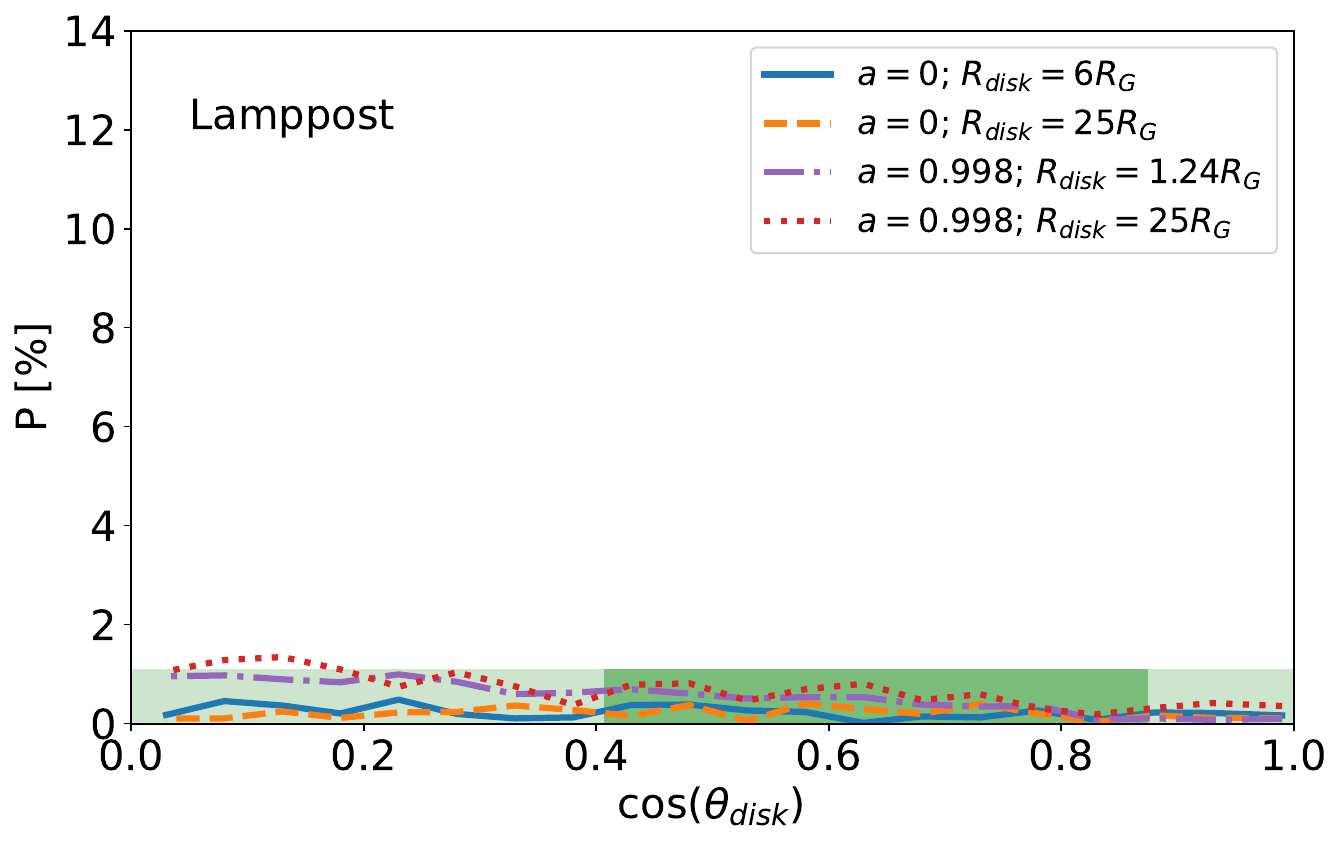}
\includegraphics[width=6cm] {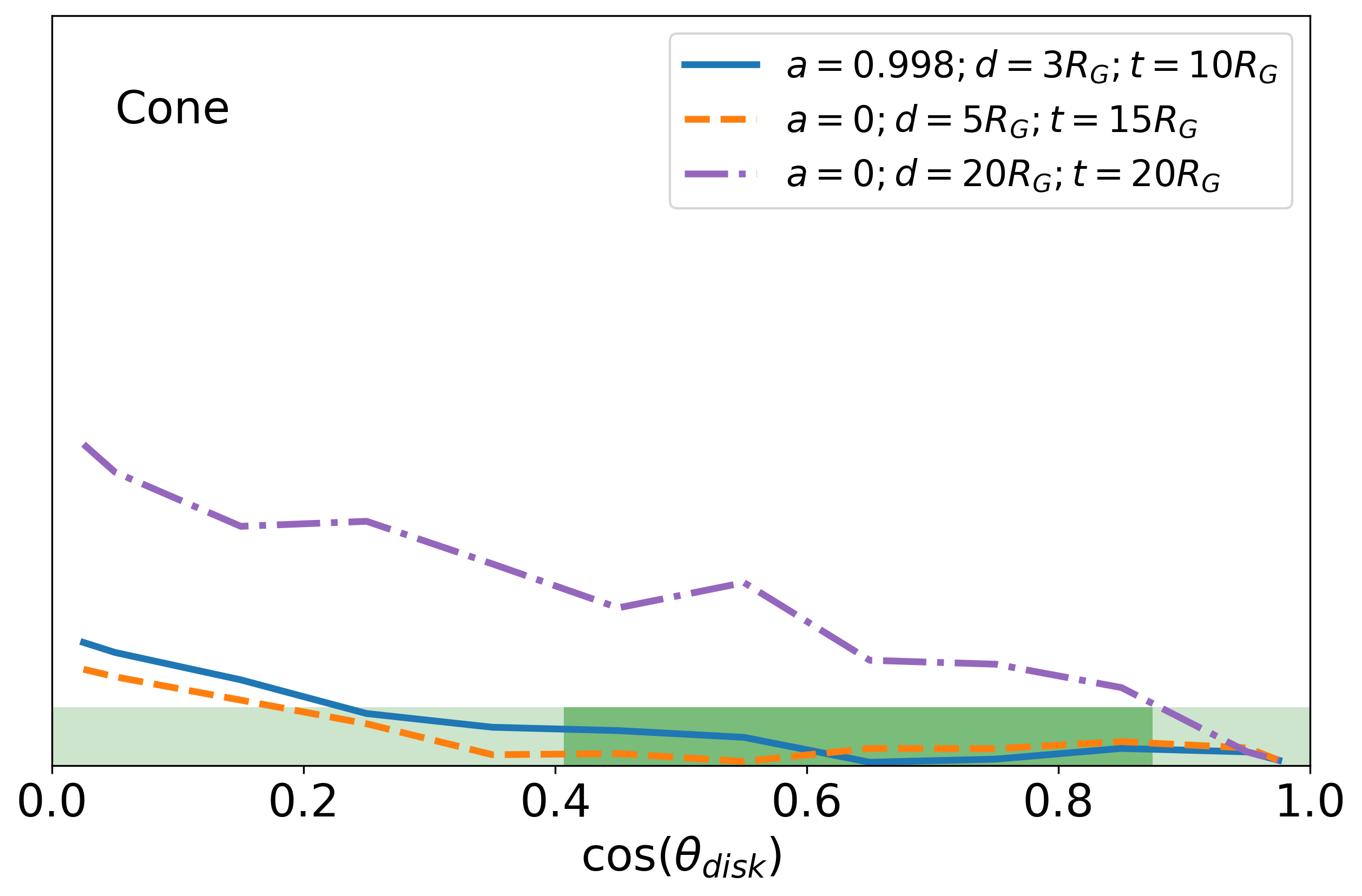}
\includegraphics[width=6cm] {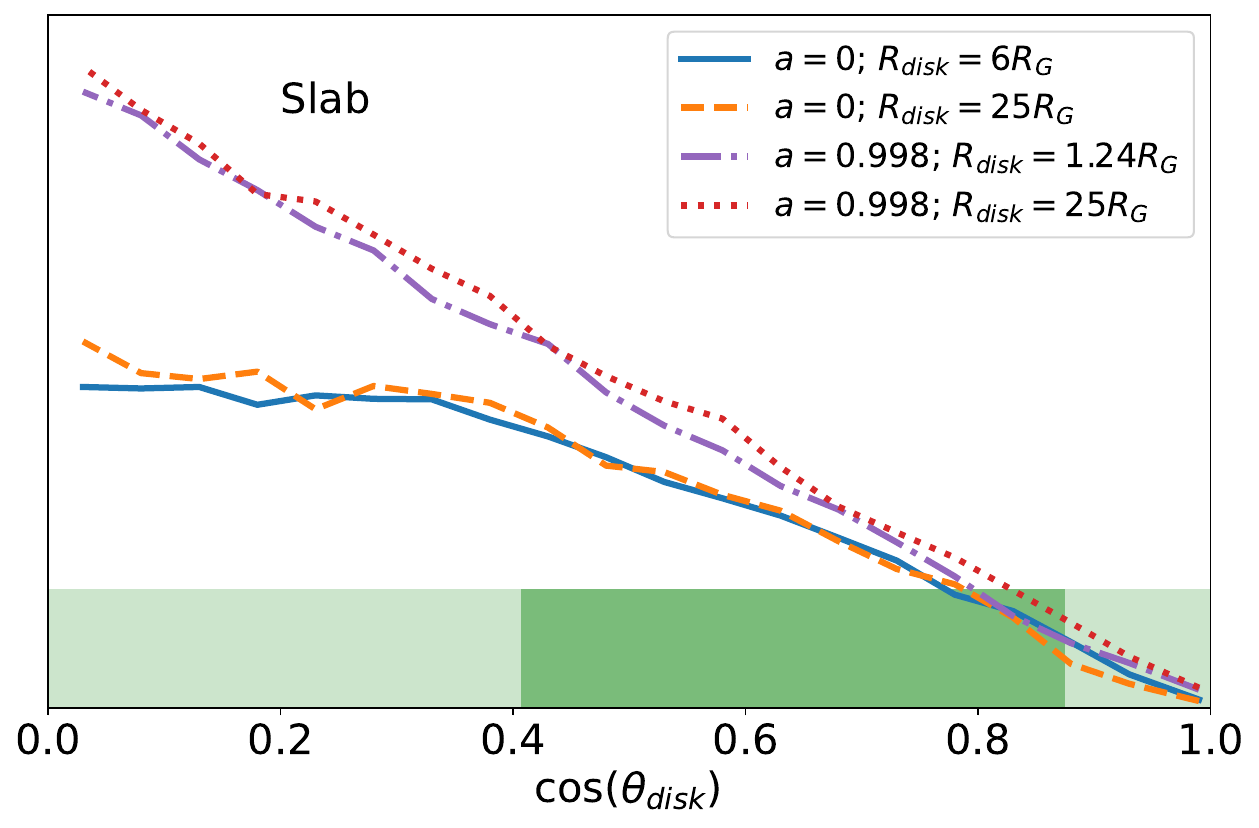}
\caption{Polarization degree from the \texttt{MONK} simulations as a function of the cosine of the system inclination angle $\cos(\theta_{\rm disk})$ in the case of a spherical lamppost model (\textit{left panel}), a conical outflows model (\textit{middle panel}) and a slab corona model (\textit{right panel}). $\cos(\theta_{\rm disk})=0$ and $\cos(\theta_{\rm disk})=1$ represent the edge-on and face-on views of the source, respectively. The green regions represent the allowed values of the polarization degree: pale green represents the constraint on $P$ found with the broad-band spectro-polarimetric analysis in Sect.~\ref{Results:Broadband_pol}, while saturated green represents the combination of the constraints on $P$ and the system inclination (Sect. \ref{Discussion:orientation}). Since the lamppost and the cone result in $\theta$ perpendicular to the accretion disk axis, the resulting polarization constraint is set to $P<1.1\%$ in the left and middle panels. Conversely, the slab results in $\theta$ parallel to the disk. In that case $P<2.4\%$.}
\label{Fig:monk_1}
\end{figure*}

\begin{figure*}
\includegraphics[width=18.75cm] {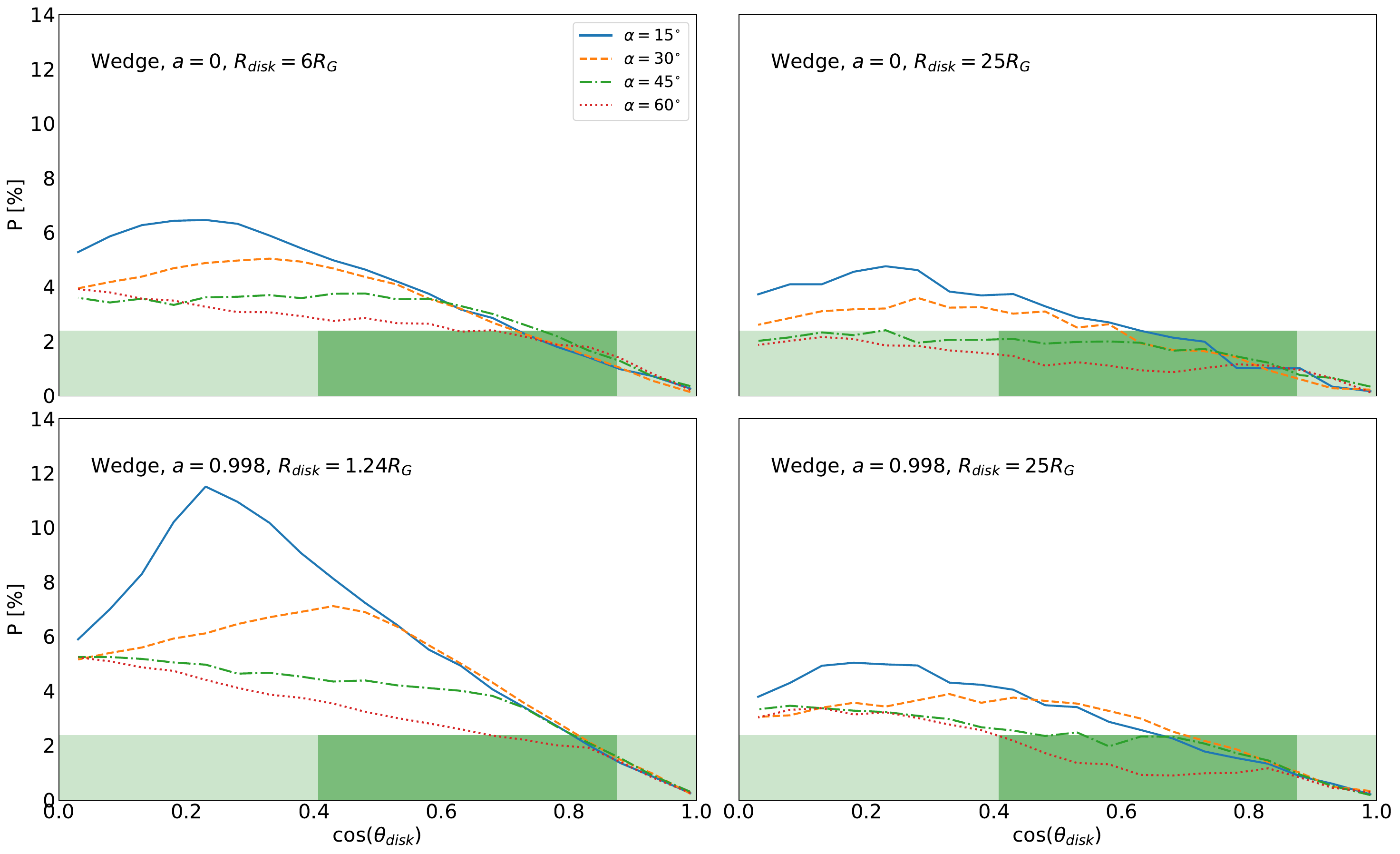}
\caption{Same as in Fig.~\ref{Fig:monk_1} but solely for the wedge corona model. This geometry results, as the slab, in $\theta$ parallel to the disk, leading to $P<2.4\%$.}
\label{Fig:monk_2}
\end{figure*}

In Figs.~\ref{Fig:monk_1} and \ref{Fig:monk_2}, we show the polarization degree $P$ as a function of the cosine of the system inclination angle, $\cos(\theta_{\rm disk})$. Since all the models are axisymmetric, $P$ is lowest for $\cos(\theta_{\rm disk})=1$ (face-on view) and highest for $\cos(\theta_{\rm disk})=0$ (edge-on view). The green regions represent the allowed values of the polarization degree: pale green indicates the constraint on $P$ derived from the broad-band spectro-polarimetric analysis in Sect.~\ref{Results:Broadband_pol}, while saturated green represents the combination of the constraints on $P$ and on the inclination of the system ($29\degree<\theta_{\textnormal{disk}}<66\degree$) from a conservative combination of estimates in various wavebands (see Sect. \ref{Discussion:orientation}).
Comparing our simulations with the data analysis results, we found that the lamppost model is compatible for all the tested scenarios, similar to the conical corona configuration, except for the case with $d=20$~R$_{\textnormal{G}}$ and $t=20$~R$_{\textnormal{G}}$. Both models predict polarization angles perpendicular to the disk axis, with stringent upper limits ($P \le 1.1\%$). Also the slab geometry yields compatible to the constraints polarization degrees, while the wedge model is compatible in the majority of the cases. These two models predict polarization angles parallel to the disk axis, and their effective polarization upper limit is only $2.4\%$. Summarizing: the combination of the inclination estimates and the upper limits on the polarization fraction lead to the inability to distinguish between different coronal models.

\section{Discussing our X-ray results through the prism of near-ultraviolet to near-infrared (spectro)polarimetry}
\label{Discussion}

Taken at face value, our X-ray polarimetric results either disfavor equatorial (slab or wedge) coronal geometries if the inclination of MCG–5–23–16 is large, as previously estimated, or indicate that the orientation of the AGN core is relatively low — in contradiction with its type~1.9 classification. To break this degeneracy, we took advantage of an unpublished optical spectro-polarimetric observation of MCG–5–23–16 obtained with the FOcal Reducer/low dispersion Spectrograph~2 (FORS2) mounted on the Very Large Telescope (\textit{VLT}).

\subsection{Data reduction}
\label{Discussion:data}

The data were acquired on May 12, 2019 at the Unit 1 (Antu) telescope, during the Run/Programme~ID 0101.B-0530(A). A Wollaston prism splits the incoming light into two orthogonally polarized beams separated by 22" on the charge-coupled device (CCD). To measure the normalized Stokes parameters $q$ and $u$, four 200 seconds-long exposures were taken with the half-wave plate rotated to 0$^\circ$, 22.5$^\circ$, 45$^\circ$, and 67.5$^\circ$, minimizing instrumental polarization. Spectra were recorded with the 300V grism and GG435 filter, covering 4600–8650~\AA. A 1" slit provided a resolving power of $R \approx 440$ at 5849~\AA, sufficient to sample broad emission lines. The slit was oriented along the parallactic angle. CCD pixels were binned $2\times2$ (0.25" per pixel). The total exposure time was 800 seconds, the airmass was constant during the observation (around 1.04) and the seeing was about 0.95". Polarized and unpolarized standards \citep{Fossati2007} were observed: Vela1-95 (Apr 1), Hiltner 652 (Jul 26, Aug 28), HD~42078 (Apr 1), and WD~1620-391 (Aug 28). Instrumental polarization was below $0.1\%$.

Raw frames were cleaned of cosmic rays with the Python implementation of \texttt{lacosmic} \citep{Dokkum2001,Dokkum2012}. Reduction used the European Southern Observatory (ESO) FORS2 pipeline \citep{Izzo2019}, yielding flat-fielded, rectified, wavelength-calibrated spectra. One-dimensional spectra were extracted with a 1.75" (7-pixel) aperture, minimizing host contamination while retaining flux. Sky spectra were taken from adjacent MOS strips and subtracted. Stokes parameters were computed from the ordinary and extraordinary beams following the FORS2 manual\footnote{\url{https://www.eso.org/sci.html}}, with correction for the half-wave plate chromatic dependence. Fluxes were corrected for atmospheric extinction and calibrated with a master response curve. Uncertainties were derived by propagating photon and readout noise. The polarization degree $P$ and angle $\theta$ were obtained as for the X-ray polarization measurements presented earlier in this paper.

\subsection{Analysis}
\label{Discussion:analysis}

\begin{figure*}
    \centering
    \includegraphics[width=\textwidth]{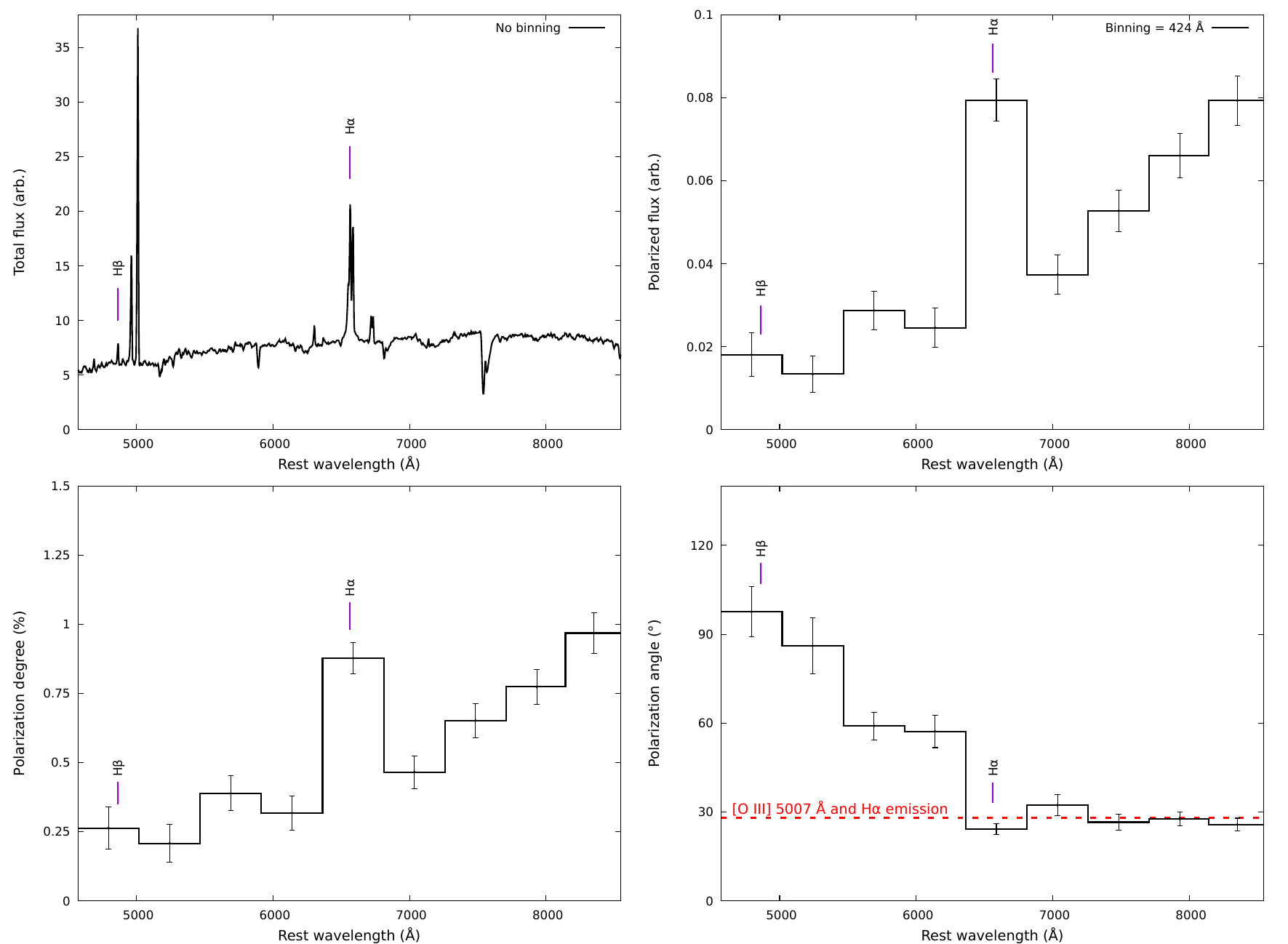}
    \caption{\textit{VLT}/FORS2 spectro-polarimetry of MCG-5-23-16. The top-left figure shows the total flux spectrum, corrected for the instrumental response (in arbitrary units). The top-right panel shows the polarized flux, that is the multiplication of the total flux with the polarization degree. The bottom-left panel presents the linear polarization degree while the bottom-right panel shows the polarization position angle. Except for the total flux panel, spectra were rebinned to 137 consecutive pixels (424.8\AA) to achieve a polarization signal-to-noise ratio of at least 3 per bin. Observational errors are indicated for each spectral bin. The newly estimated position angle of the extended [O III] and H$\alpha$ emission line region observed by \citep{Ferruit2000} is shown with a red dashed line in the polarization angle panel.}
    \label{Fig:Spectropola_binned}
\end{figure*}

We present the results of this observation in Fig.~\ref{Fig:Spectropola_binned}. The total flux spectrum is shown at native spectral resolution but, because the polarization was too noisy (see Fig.~\ref{Fig:Spectropola_unbinned} in appendices for the unbinned version), we binned the polarized spectra so that each spectral bin has a polarization signal-to-noise ratio of at least 3. The total flux spectrum shows only narrow emission lines on top of a reddened continuum, a spectral shape already known for decades \citep[see, e.g.][]{Lumsden2004}. We fitted the main Balmer lines and found a full width half maximum (FWHM) of about 410 km~s$^{-1}$ for H$\alpha$ and about 330 km~s$^{-1}$ for H$\beta$. The unbinned polarized flux, however, shows a larger H$\alpha$ line, with a FWHM of $\sim$ 2250 km~s$^{-1}$, but no detectable H$\beta$ emission, see Fig.~\ref{Fig:Spectropola_unbinned}. The H$\alpha$ FWHM we find is comparable to the $\sim$ 2150 km~s$^{-1}$ quoted in \citet{Blanco1990} for Br$\gamma$, indicating that both the emission lines share a common physical origin. The polarized continuum is significantly redder than the total flux, indicating that the polarized light is affected by substantial dust extinction, which preferentially suppresses the blue wavelengths. The polarization degree (Fig.~\ref{Fig:Spectropola_binned}, bottom-left panel) shows a similar trend, with $P$ of the order of 0.25\% around 5000~\AA, a peak at the location of the H$\alpha$ line and a rise towards 9000~\AA, reaching up to 1\%. Remarkably, the chromaticity of $P$ is closely coupled to a rotation of the polarization angle from about 100$^\circ$ in the blue band to $\sim$ 30$^\circ$ longward of H$\alpha$, where $\theta$ stabilizes up to the end of the red band. The combination of those two independent quantities points toward a change of polarization mechanism between the two ends of the spectrum. 

To solve this riddle, we first checked if our data is contaminated by interstellar polarization (ISP), that is to say the effect of Milky Way's dust and magnetized interstellar medium. To do so, we used the compilation of optical starlight polarization catalogs presented by \citet{Panopoulou2025} and selected the closest star with available polarimetric measurement from the compilation: GaiaDR2~3244105248118361984. This star is situated 2.27$^\circ$ away from MCG-5-23-16 and its polarization is $P$ = 0.25\% $\pm$ 0.1\% at $\theta$ = 39.1$^\circ$ $\pm$ 12.8$^\circ$. While the measured red band polarization observed in MCG-5-23-16 is higher than the ISP value by 0.72\% $\pm$ 0.12\% -- which corresponds to a significance of about 5.8$\sigma$ --, the blue band shows a similar polarization level, but with a different polarization angle. This suggests that ISP may contribute to the observed polarization in the blue band, with dichroic absorption dominating the production of polarization, overlaid with a second component that drives the rotation of $\theta$. 

\begin{figure}
    \centering
    \includegraphics[width=1\columnwidth,trim={0.5cm 0cm 0cm 0cm}]{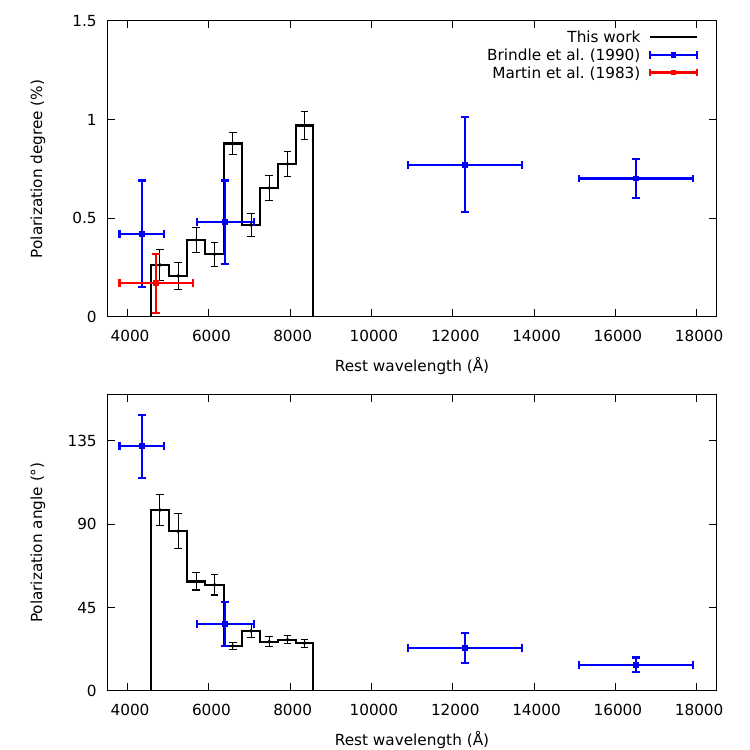}
    \caption{Our spectro-polarimetric results, compared to the broad-band measurements of \citet{Martin1983} and \citet{Brindle1990}, in red and blue, respectively. The polarization angle is unconstrained in the single measurement of \citet{Martin1983}.}
    \label{Fig:Old_measurements}
\end{figure}

We then compared our data with previous polarimetric measurements to see if they agree and whether the chromaticity in $P$ and $\theta$ extends to shorter and longer wavelengths. There are very few past polarimetric observations of this source in the literature: two papers reporting broad-band polarimetry \citep{Martin1983,Brindle1990} and one paper with spectro-polarimetry \citep{Lumsden2004}. The latter data were of lower signal-to-noise ratio and add little new information so, on Fig.~\ref{Fig:Old_measurements}, we only superimposed the broad-band U, R, J and H polarimetric points of \citet{Brindle1990} and the Coming 4-96 filter (3800-5600~\AA) measurement of \citet{Martin1983} to our data, keeping in mind that they were obtained with larger apertures (between 4" and 6") and different observing conditions than ours. Yet, the results strongly agree with ours, both in terms of $P$ and $\theta$. The polarization of this source appears to have remained unchanged for decades. More importantly, the U-band polarimetric point of \citet{Brindle1990} confirms the rotation of the polarization angle at short wavelengths, with a value of 132$^\circ$ $\pm$ 17$^\circ$, i.e. orthogonal to the polarization angle in the red band. The J and H-band polarization angles confirm that $\theta$ remains stable, at least up to 1.8~$\mu$m. The polarization degree in the infrared appears to stabilize too, at about 0.8\%, compatible with our red band point, within the error bars. 

With all this information in hand, we have one more element to examine to determine what is happening with the wavelength-dependent polarization of MCG-5-23-16 in the optical and the non-detection of its X-ray polarized counterpart: the alignment of the polarization position angle with the radio jet or AGN outflows axes. This is a crucial test that was revealed by \citet{Antonucci1983} for radio galaxies first, then for Seyfert galaxies, and finally extended to all non-blazing AGNs \citep{Antonucci1993}. Comparing the two provides insight into the global AGN orientation and helps identify the location (equatorial vs. polar) of the regions responsible for the observed polarization. The position angle of the radio jet structure in MCG-5-23-16 is subject to debate, as it is only marginally resolved on the 1.4~GHz Karl G. Jansky Very Large Array (VLA) images taken by \citet{Mundell2009}, see also \citet{Orienti2010}. The estimated position angle of the radio structure is then approximately -11$^\circ$, which is neither parallel nor perpendicular to the optical polarization angle, whether in the blue or red bands. In the absence of a clear radio structure, it is common to look at the position angle of the AGN ionization axes, highlighted in the optical and infrared bands by their emission line outflows. Past observations revealed that the [O~III] emission in MCG-5-23-16 is elongated on either side of the nucleus at a position angle of 40$^\circ$, uncorrelated to the marginally extended radio emission \citep{Ferruit2000}. If we take this value as the reference angle to which compare our polarization angle (26$^\circ$ $\pm$ 2$^\circ$ at 8424~\AA), the two appear nearly aligned, but differ by $\sim$ 14$^\circ$, corresponding to a $>$ 6$\sigma$ discrepancy (accounting for the modulo-180$^\circ$ ambiguity).

\begin{figure}
    \centering
    \includegraphics[width=1\columnwidth]{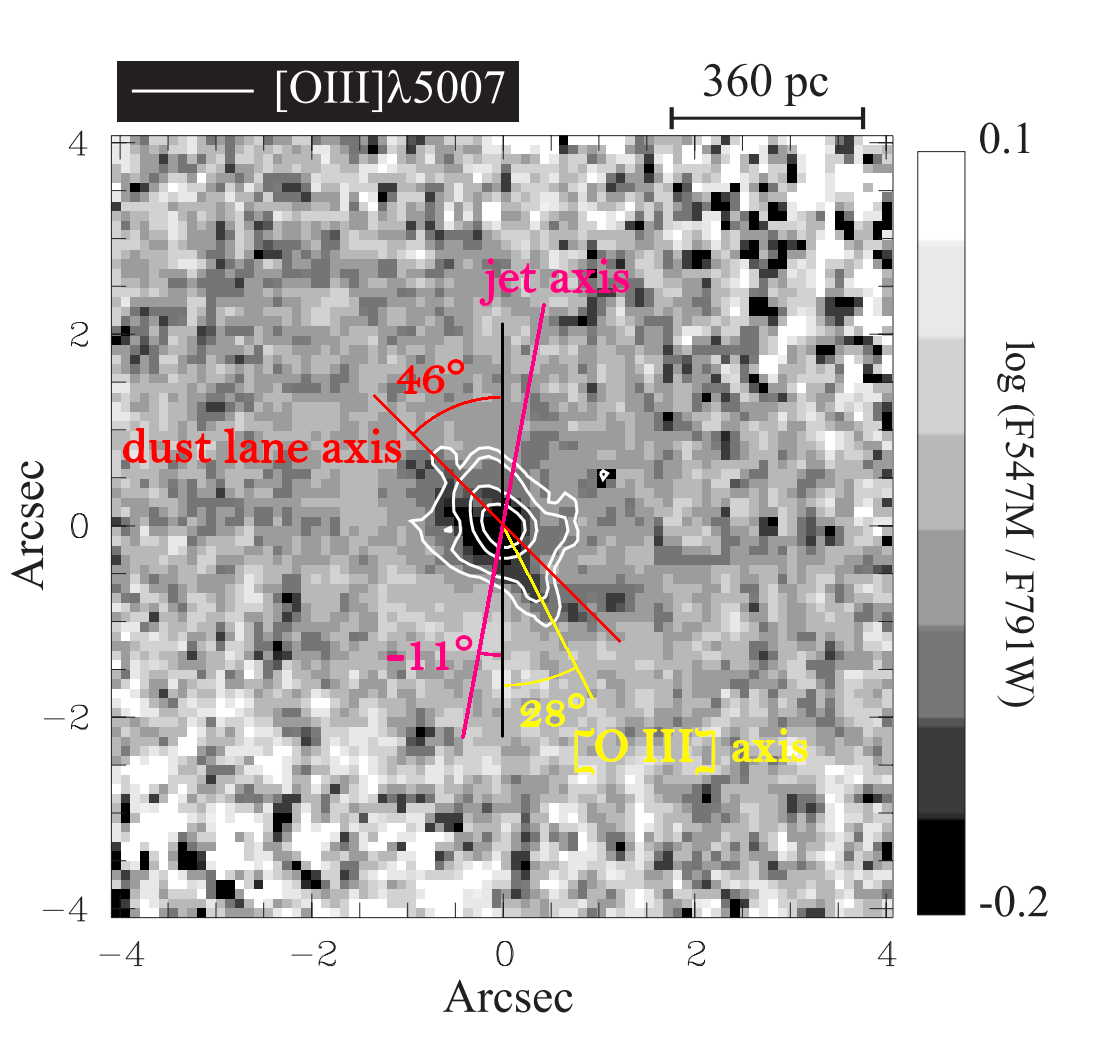}
    \caption{Color map of MCG-5-23-16 taken with the HST/WFPC2, extracted from the publication of \citet{Ferruit2000}, see their Fig.~8. North is up and east to the left. The color map was constructed by dividing the F547M continuum image by the F791W one and taking the logarithm of the result. Darker regions [lower values of log (F547M/F791W)] correspond to redder areas. Selected isophotes (8, 16, 64, and 256 $\times$ 10$^{-15}$ ergs~s$^{-1}$~cm$^{-2}$~arcsec$^{-2}$) of the [O III] $\lambda$5007 image have been superposed on the image. The gray scales are linear and the pixel size is 0.0996~arcsec$^{-2}$. In pink, we traced on top of the \citet{Ferruit2000}'s image the position angle of the radio jet axis, marginally measured by \citet{Mundell2009} and \citet{Orienti2010}. In red, we show our measured value of the dust lane position angle extracted from \citet{Prieto2014}. Finally, in yellow, we traced our new estimation of the position angle of the maximal extension of the [O III] $\lambda$5007 emission.}
    \label{Fig:Angles}
\end{figure}

To better understand the underlying situation, we look more carefully at the extension and position angle of the outflows, as revealed by their line emission in \citet{Ferruit2000}. In Fig.~\ref{Fig:Angles}, we digitized and reproduced their color map of MCG-5-23-16, taken with the Wide Field and Planetary Camera 2 (WFPC2) onboard the Hubble Space Telescope (HST), with their [O III] $\lambda$5007 image superposed on the image. We stress that we did not re-reduce the data, we simply extracted their figure. In this image, we traced in pink the position angle of the radio jet axis, originally estimated by \citet{Mundell2009}, for reference. In yellow, we measured the position angle of the [O III] region by identifying the direction of maximum spatial extension of the forbidden emission above a given surface-brightness threshold. The position angle corresponds to the vector joining the nucleus to the most distant pixel belonging to the contiguous emission region in the south-west quadrant. Our measurement\footnote{Although the measurement depends on the adopted surface-brightness threshold, we verified that varying this threshold within a reasonable range does not significantly affect the derived position angle. The position angle remains stable within 4$^\circ$ for all tested thresholds, indicating that the result is robust against the specific choice of contour level.} yields a position angle of 28$^\circ$ $\pm$ 4$^\circ$, a value rather different than what was proposed by \citet{Ferruit2000}. The value reported by the authors (40$^\circ$) rather concerns a more marginal extension in the opposite quadrant -- the north-east one -- which could be distorted by the presence of an emission blob, as potentially suggested by the shape of the isophotes. The position angle we measure is  43$^\circ$ $\pm$ 6$^\circ$ for this extension. The problem is that this value comes into conflict with the observed optically thick dust lane that is seen in projection crossing very close to the AGN center, at a position angle of about 46$^\circ$ \citep{Prieto2014}, reported in red in our Fig.~\ref{Fig:Angles}. This dust lane -- that is part of a much larger kpc-scale structure \citep{Prieto2014} -- is either an isolated filament or the edge-on disk of the host galaxy itself \citep{Esparza2025}, but what is certain is that the [O III] outflow in this quadrant is most certainly partially hidden by the veil of dust, so its position angle must be regarded as highly putative. For this reason, we therefore focus on the less dust-affected south-west quadrant.

By doing so, and accounting for the dust lane, the pieces finally fit together. The position angle of the ionization axis of the AGN becomes parallel to the near-infrared and infrared polarization, as shown in red in the bottom-right panel of our Fig.~\ref{Fig:Spectropola_binned}, and perpendicular in the blue band. In the infrared, dust extinction from the filament is weaker, allowing us to directly observe the core emission of MCG-5-23-16, including the broad Pa$\beta$ emission from the broad line region (BLR), see \citet{Goodrich1994}. As we move towards the optical band, dust extinction increases (as highlighted by the shape of the polarized flux, see Fig.~\ref{Fig:Spectropola_binned}, top-right panel) and photons can only reach the observer after scattering off the polar outflows. This is the reason why we observe H$\alpha$ in polarized light only. Since the optical core photons are scattered perpendicularly in the narrow line region (NLR), the polarization angle rotates, reaching an orthogonal value shortward of H$\alpha$. The fact that $P$ decreases in the blue part of our spectrum is due to a combination of starlight and ISP that cancels the small polarized flux we get from core AGN photons that have scattered off the NLR, hence the non-detection of broad H$\beta$ line (further compounded by the lower signal-to-noise ratio in the blue). Such a variety of detection and non-detection of broad lines in the blue, optical, and infrared bands arises because the true cause of the nucleus obscuration is not the usual optically-thick, compact, circumnuclear reservoir of dust and gas that we usually call the torus, but the less-dense, distant dust lane bisecting the galaxy. This means that MCG–5-23-16 is not intrinsically a type-1.9/2 AGN, but a type-1 nucleus seen through a foreground dust lane, unrelated to the AGN itself. This also explains why the amount of hydrogen column density towards MCG-5-23-16 is so low (n$_{\rm H}$ $\approx$ 10$^{22}$ at~cm$^{-2}$, see Sect.~\ref{Results:Broadband}), why it did not show any significant variation in decades (which is consistent with an obscurer at kilo-parsec scale, \citealt{Serafinelli2023}), and why MCG-05-23-16 continues to elude X-ray polarimetric detection. Its accretion structure, hot corona, and BLR are seen almost face-on, resulting in a very low expected X-ray polarization degree, even from equatorial models (see Fig.~\ref{Fig:monk_2} and \citealt{Tagliacozzo2025}).

\subsection{The true orientation of MCG-05-23-16's nucleus}
\label{Discussion:orientation}

The idea that obscuration in some Seyfert galaxies arises primarily from a host-galaxy dust lane rather than a compact torus has been argued in the literature (e.g., \citealt{Maiolino1995,Matt2000,Bianchi2012}). Our results provide compelling evidence that this is also the case for MCG-05-23-16 \citep{Prieto2014}. The Seyfert galaxy is, in fact, a type-1 AGN, unintentionally disguised as a type-1.9/2 AGN. This also explains why it has been so difficult to estimate its core orientation. 

\citet{Turner1998}, using ASCA data, fitted the iron K$\alpha$ line on top of a single power-law continuum in a relativistic framework and estimated the inclination of the accretion disk in MCG-05-23-16 to be 33$^{+11}_{-4}\degree$. \citet{Weaver1997} found a larger inclination using the same methodology (about 65$^\circ$), likely caused by the assumptions made in the fit. \citet{Reeves2007} modeled the Suzaku X-ray spectrum of this source using a combination of reflection off distant matter and the accretion disk, and determined that the most probable inclination of MCG-05-23-16 is $\sim$ 50$^\circ$. Simultaneous deep XMM-Newton and Chandra observations of MCG-5-23-16 lead \citet{Braito2007} to achieve a time-averaged spectral analysis of the iron K-shell complex, well described by an emission line originating from an accretion disk viewed with an inclination angle of about 40$^\circ$. Taking advantage of the first contemporaneous X-ray observation of this AGN by XMM-Newton and NuSTAR, \citet{Serafinelli2023} analyzed the shape of the broad iron line and determined an inclination of 41$^{+9}_{-10}\degree$. Using a different technique, \citet{Fuller2016} compiled infrared fluxes of MCG-05-23-16 and used numerical torus models together with a Bayesian approach to fit its infrared nuclear spectral energy distribution. They constrained the torus inclination to 
62$^{+4}_{-5}\degree$. Even in our own broadband analysis (see Sect.~\ref{Results:Broadband}), $\theta_{\textnormal{incl}}$ is unbound unless constrained before the fit.

We see that, depending on the waveband, model and assumptions, the estimated inclination of MCG-05-23-16 ranges between 29$^\circ$ and 66$^\circ$, based on the quoted papers. It has been established that determining the orientation of an AGN core is extremely tricky and that there is not a single method that gives consistent results for a large test sample \citep{Marin2016}. One possible reason behind this discrepancy is that MCG-05-23-16 was initially considered a type-1.9/2 AGN based solely on its optical and near-infrared spectra. This optical classification naturally led some authors to favor models with higher inclinations (see, e.g., \citealt{Weaver1997}). Now that it is understood that MCG-05-23-16 is in fact a type-1 AGN seen through a foreground dust lane, future modeling can be performed without this previous assumption, allowing a more reliable reassessment of its inclination.

\section{Conclusions}
\label{Conclusion}

In this paper, we report the third \textit{IXPE} pointing of MCG-05-23-16, together with new optical spectro-polarimetry obtained with the \textit{VLT}, and combined with archival near-ultraviolet, optical and near-infrared polarimetric data. No polarization was detected in the 2-8~keV energy range at an upper limit of $\le$~2.9\% (99\% confidence level). Combining the new \textit{IXPE} pointing with the past two (May and November 2022) helped reducing the upper limit to $\le$~2.5\% (99\% confidence level). Monte Carlo simulations of the X-ray corona either indicate that equatorial models (slab, wedge...) are disfavored if the AGN is highly inclined (as induced by its historical type-1.9/2 classification). However, a corona coplanar to the accretion disk can be a viable solution if the source is less inclined than what is usually thought. 

To determine which of the two scenarios is the correct one, we analyzed a 2019 \textit{VLT}/FORS2 dataset to obtain optical and near-infrared spectro-polarimetry of the source. The total flux spectrum displayed only narrow emission lines on top of a reddened continuum, consistent with a Seyfert-2/1.9 classification. However, the polarized spectrum revealed a broad H$\alpha$ in scattered light and a strong wavelength dependence of both the polarization degree and angle, with a rotation of nearly 70$^\circ$ between the blue and red bands. Comparison with historical near-ultraviolet, optical, and infrared measurements demonstrated 1) long-term stability of the polarization spectrum, pointing to a persistent mechanism, and 2) that the rotation of the polarization angle reaches 90$^\circ$ in the U-band. 

After defining a correct reference for comparing our polarization data, namely the ionization axis of the AGN, we showed that the observed polarization is best explained by a type-1 nucleus viewed through a large-scale, distant and moderately obscuring dust lane rather than by an optically-thick, compact circumnuclear torus. This reinterpretation explains why the source has such a moderate  hydrogen column density for an AGN that was supposedly so much inclined. This is in agreement with the absence of an X-ray polarimetric detection by IXPE, since the source is effectively a face-on type-1 AGN whose hot corona yields intrinsically low X-ray polarization. Our results remove MCG-05-23-16 from the sample of bona fide Seyfert-1.9 galaxies and call for a reassessment of its orientation using models free from this long-standing misclassification.

\begin{acknowledgements}
The \textit{Imaging X ray Polarimetry Explorer} ({\em IXPE}) is a joint US and Italian mission. The US contribution is supported by the National Aeronautics and Space Administration (NASA) and led and managed by its Marshall Space Flight Center (MSFC), with industry partner Ball Aerospace (contract NNM15AA18C). The Italian contribution is supported by the Italian Space Agency (Agenzia Spaziale Italiana, ASI) through contract ASI-OHBI-2022-13-I.0, agreements ASI-INAF-2022-19-HH.0 and ASI-INFN-2017.13-H0, and its Space Science Data Center (SSDC) with agreements ASI-INAF-2022-14-HH.0 and ASI-INFN 2021-43-HH.0, and by the Istituto Nazionale di Astrofisica (INAF) and the Istituto Nazionale di Fisica Nucleare (INFN) in Italy. This research used data products provided by the {\em IXPE} Team (MSFC, SSDC, INAF, and INFN) and distributed with additional software tools by the High-Energy Astrophysics Science Archive Research Center (HEASARC), at NASA Goddard Space Flight Center (GSFC). This work made use of data supplied by the UK Swift Science Data Centre at the University of Leicester. DH is research director at the F.R.S-FNRS, Belgium. V.E.G. acknowledges funding under NASA contract 80NSSC24K1403. POP acknowledges financial support from the french national space agency (CNES) and the CNRS ATPEM. R.S. acknowledges funding from the CAS-ANID grant number CAS220016.

J.S. thanks GACR project 21-06825X for the support. D.H. is F.R.S-FNRS research director (Belgium).
        
\end{acknowledgements}

\bibliographystyle{aa} 
\bibliography{Bibliography} 

\begin{appendix}
\onecolumn

\section{X-ray observations}
\label{X-ray_obs}

\begin{figure*}
\includegraphics[width=1.\columnwidth] {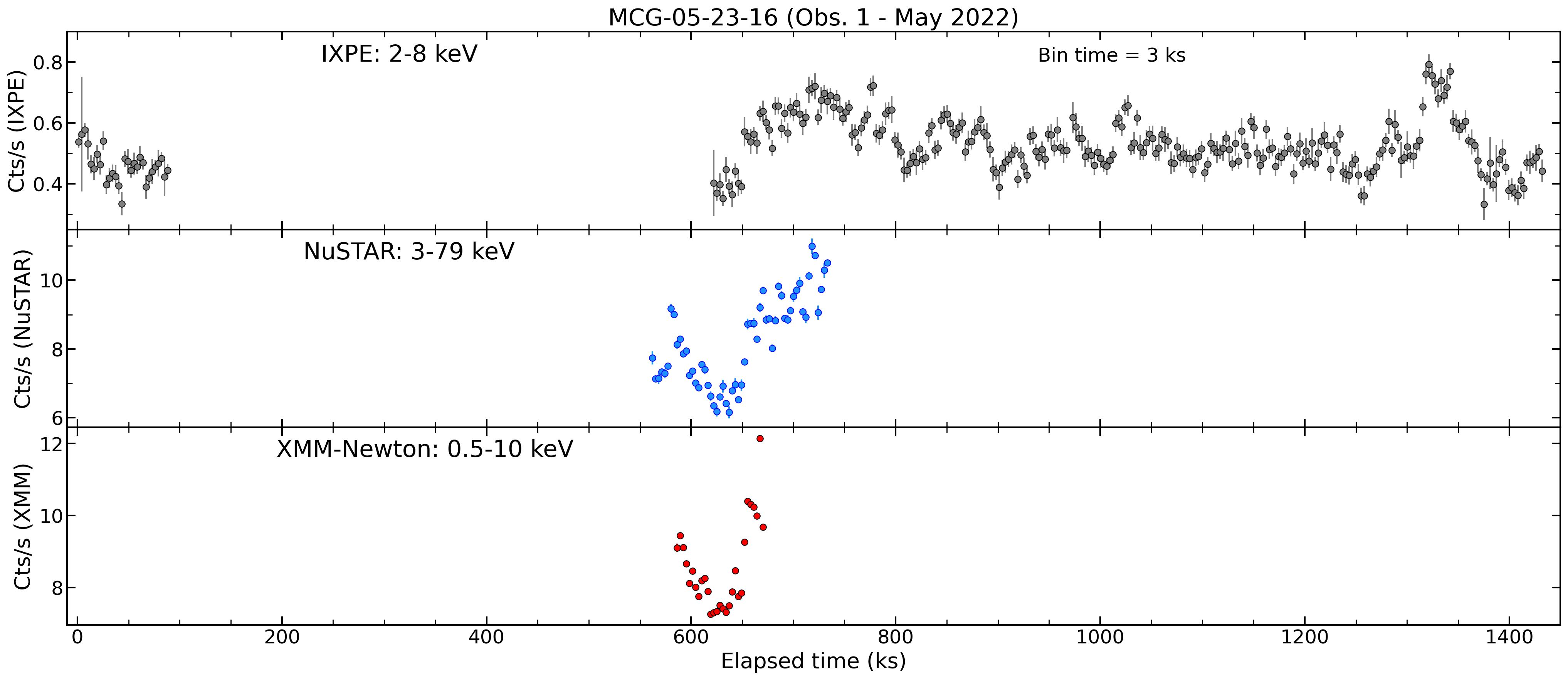}
\includegraphics[width=1.\columnwidth] {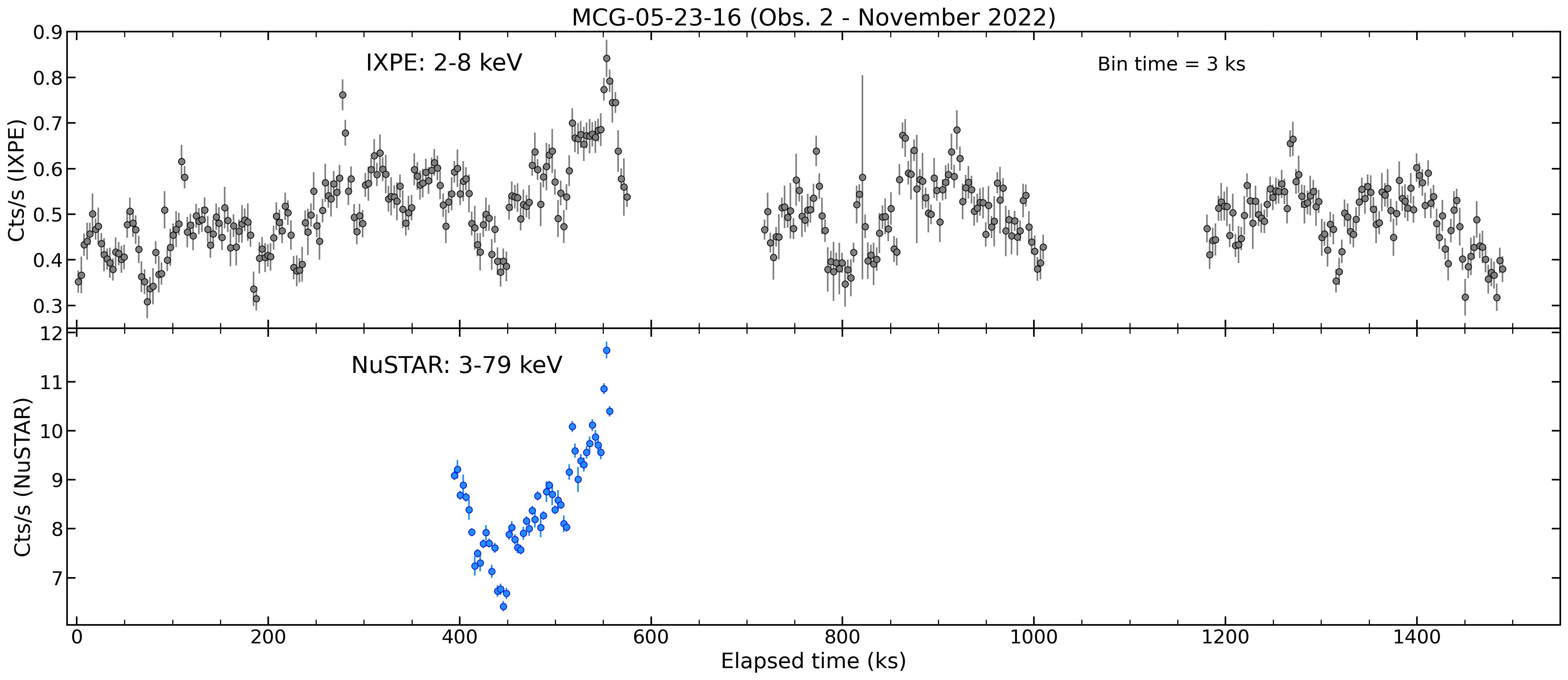}
\includegraphics[width=1.\columnwidth] {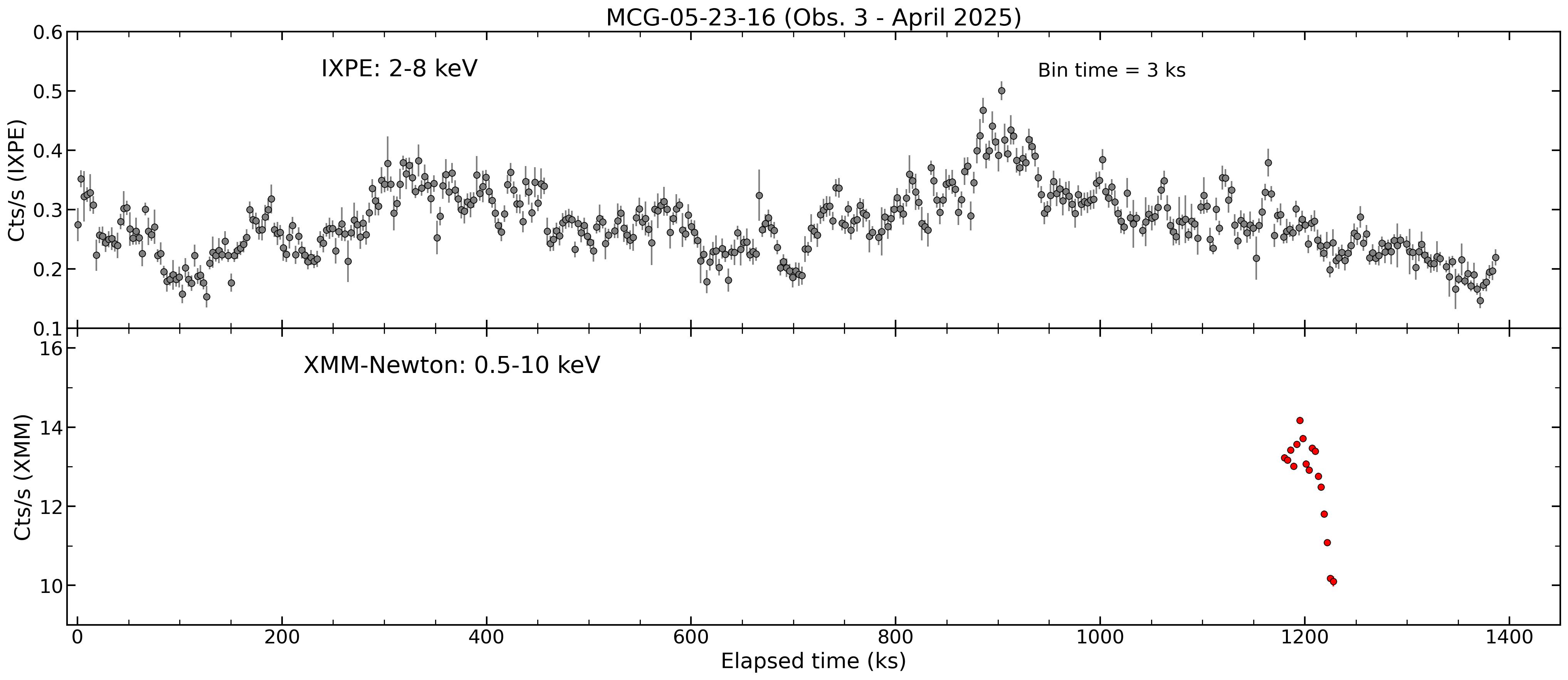}
\caption{\textit{IXPE}, {\it NuSTAR} and \textit{XMM-Newton} light curves of the three observing campaigns of MCG-05-23-16 are shown. Except for the third {\it IXPE} observation, where we only had DU1 and DU3 available for the analysis, data counts from DU1, DU2 and DU3 on board of \textit{IXPE} and from FPMA/A and FPMA/B on board of {\it NuSTAR} have been summed. The full energy bands of the three satellites have been used and we adopted a 3 ks time binning.}\label{lc}
\end{figure*}

In Fig. \ref{lc} we show the light curves of the three X-ray observing campaigns of MCG-05-23-16 used in this work.

\section{Optical spectro-polarimetry}
\label{Spectropolarimetry_unbinned}

\begin{figure*}
    \centering
    \includegraphics[width=\textwidth]{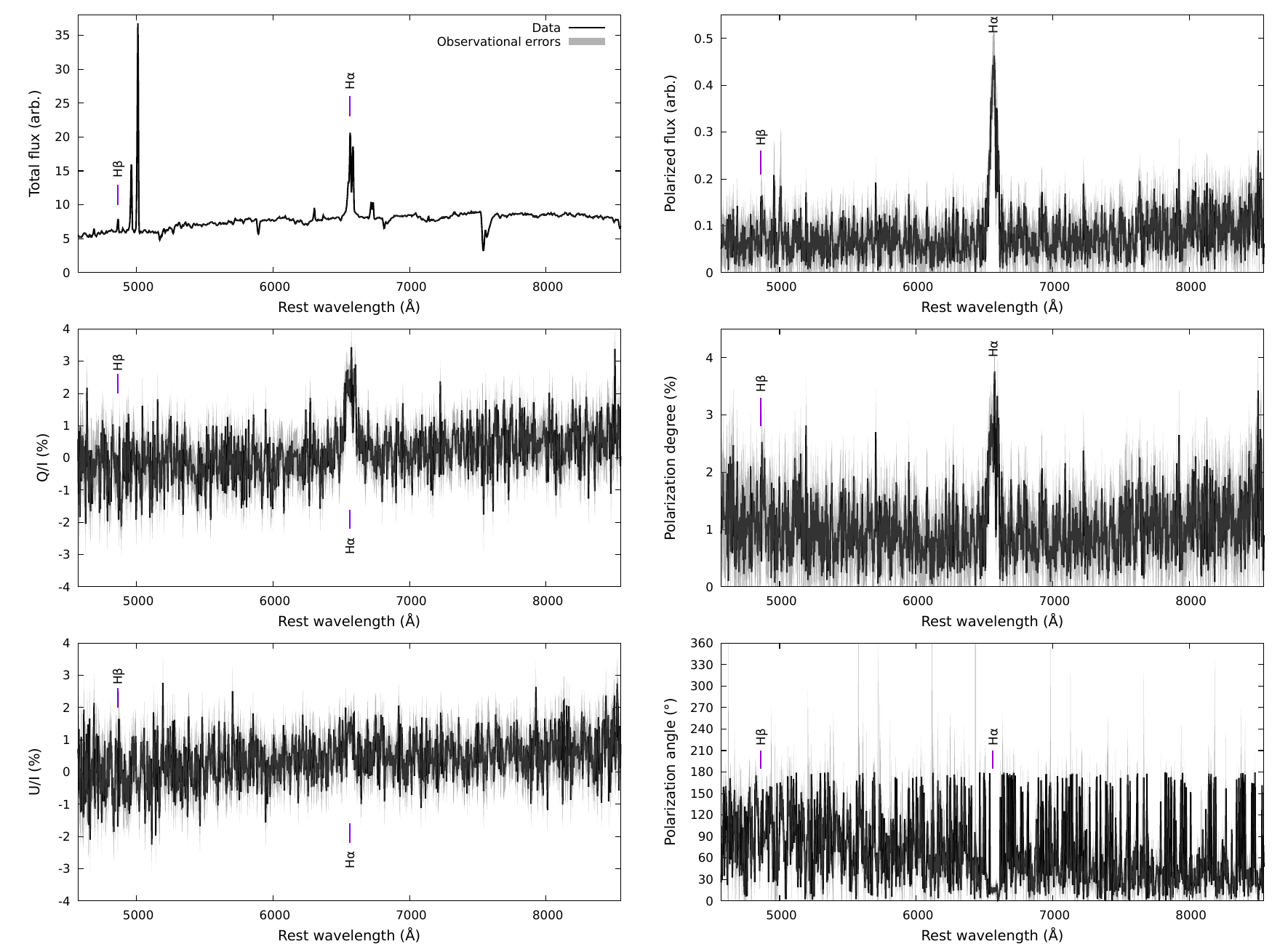}
    \caption{\textit{VLT}/FORS2 spectro-polarimetry of MCG-05-23-16. The top-left figure shows the total flux spectrum, corrected for the instrumental response (in arbitrary units). The middle- and bottom-left figures present the Q/I and U/I normalized Stokes parameters. The top-right panel shows the polarized flux, that is the multiplication of the total flux with the polarization degree. The middle-right panel presents the linear polarization degree while the bottom-right panel shows the polarization position angle. Spectra are shown at native spectral resolution. Observational errors are indicated in gray.}
    \label{Fig:Spectropola_unbinned}
\end{figure*}

In Fig.~\ref{Fig:Spectropola_unbinned}, we present the unbinned total and polarized fluxes, the Q/I and U/I normalized Stokes parameters, and the polarization degree and angle of MCG-05-23-16 obtained with the grism 300V and the order-sorting filter GG435 with the \textit{VLT}/FORS2. In contrast to Fig.~\ref{Fig:Spectropola_binned}, the polarized signal was not rebinned for the polarized counterpart. We thus observe the poor signal-to-noise ratio of the observation in polarization, that is particularly visible in the polarization angle panel. However, despite the statistical fluctuations, the broad H$\alpha$ emission line is clearly visible in the Q/I, polarized flux, and polarization degree spectra. The width of the line is larger than in the optical, as already mentioned and measured by \citet{Lumsden2004}. This observation confirms that the broad H$\alpha$ line is indeed revealed in the polarized flux in MCG-05-23-16. No broad H$\beta$ emission line is detected in the polarized flux spectrum, but it does not indicate that the line is absent. It could very well be hidden by bad statistics.

For completeness, we also provide the polarization values of each of the spectral bins presented in Fig.~\ref{Fig:Spectropola_binned} in Tab.~\ref{Tab:VLT_binned_P}.

\begin{table}
\centering
\caption{Optical polarization in the binned VLT/FORS2 spectrum.}
\begin{tabular}{ccc}
\hline
\textbf{Central bin (Angs.)} & \textbf{$P$ (\%)} & \textbf{$\theta$ ($^\circ$)}  \\ 
\hline
4835.45	$\pm$	226.05	&	0.26	$\pm$	0.08	&	97.6	$\pm$	8.4	\\
5287.55	$\pm$	226.05	&	0.21	$\pm$	0.07	&	86.1	$\pm$	9.5	\\
5739.65	$\pm$	226.05	&	0.39	$\pm$	0.06	&	59.0	$\pm$	4.6	\\
6191.75	$\pm$	226.05	&	0.32	$\pm$	0.06	&	57.2	$\pm$	5.5	\\
6643.85	$\pm$	226.05	&	0.88	$\pm$	0.06	&	24.2	$\pm$	1.8	\\
7095.95	$\pm$	226.05	&	0.47	$\pm$	0.06	&	32.3	$\pm$	3.6	\\
7548.05	$\pm$	226.05	&	0.65	$\pm$	0.06	&	26.6	$\pm$	2.7	\\
8000.15	$\pm$	226.05	&	0.77	$\pm$	0.06	&	27.6	$\pm$	2.3	\\
8424.20	$\pm$	226.05	&	0.97	$\pm$	0.07	&	25.8	$\pm$	2.1	\\
\hline
\end{tabular}
\label{Tab:VLT_binned_P}
\end{table}

\end{appendix}

\end{document}